\documentclass[journal,twoside,web]{IEEEtran}
\usepackage{amsmath,amsfonts}
\usepackage{algorithmic}
\usepackage{algorithm}
\usepackage{array}
\usepackage[caption=false,font=normalsize,labelfont=sf,textfont=sf]{subfig}
\usepackage{textcomp}
\usepackage{stfloats}
\usepackage{url}
\usepackage{verbatim}
\usepackage{graphicx}
\usepackage{algorithm}
\usepackage{setspace}
\usepackage{color}
\usepackage{cite}
\usepackage{booktabs}

\newtheorem{remark}{Remark}
\newtheorem{lemma}{Lemma}
\newtheorem{theorem}{Theorem}

\newtheorem{assumption}{Assumption}

\def\BibTeX{{\rm B\kern-.05em{\sc i\kern-.025em b}\kern-.08em
    T\kern-.1667em\lower.7ex\hbox{E}\kern-.125emX}}
\begin{document}
\title{Data-Based Efficient Off-Policy Stabilizing Optimal Control Algorithms for Discrete-Time Linear Systems via Damping Coefficients
\thanks{
}}
\author{Dongdong Li, Jiuxiang Dong
\thanks{Dongdong Li and Jiuxiang Dong are with the College of Information Science and Engineering, Northeastern University, Shenyang 110819, China,
the State Key Laboratory of Synthetical Automation of Process Industries, Northeastern University, Shenyang 110819, China, and
the Key Laboratory of Vibration and Control of Aero-Propulsion Systems
Ministry of Education, Northeastern University, Shenyang 110819, China. Email: lidongdongyq@163.com,  dongjiuxiang@ise.neu.edu.cn.
}}

\maketitle

\begin{abstract}
Policy iteration is one of the classical frameworks of reinforcement learning, which requires a known initial stabilizing control. However, finding the initial stabilizing control depends on the known system model. To relax this requirement and achieve model-free optimal control, in this paper, two different reinforcement learning algorithms based on policy iteration and variable damping coefficients are designed for unknown discrete-time linear systems. First, a stable artificial system is designed, and this system is gradually iterated to the original system by varying the damping coefficients. This allows the initial stabilizing control to be obtained in a finite number of iteration steps. Then, an off-policy iteration algorithm and an off-policy $\mathcal{Q}$-learning algorithm are designed to select the appropriate damping coefficients and realize data-driven. In these two algorithms, the current estimates of optimal control gain are not applied to the system to re-collect data. Moreover, they are characterized by the fast convergence of the traditional policy iteration. Finally, the proposed algorithms are validated by simulation.
\end{abstract}


\section{Introduction}
\subsection{Background}
Reinforcement learning (RL) makes optimal decision by learning and interaction between the agent's actions and the external environment \cite{sutton1998introduction,lewis2012optimal}. Unlike dynamic programming, it can overcome the curse of dimensionality and the lack of a physical model \cite{bertsekas1996dynamic}. This makes RL widely used for solving optimal control and decision problems without system models, including the optimal control problems of discrete-time (DT) \cite{hewer1971iterative,chen2022robust,jiang2019optimal,kiumarsi2017h,fan2019model,kiumarsi2014reinforcement,lewis2010reinforcement} and continuous-time (CT) systems \cite{jiang2012computational,7444144,bian2016value}.
For nonlinear systems, the solution of the Hamilton-Jacobi-Bellman equation is the optimal control solution, while for linear systems, the solution of the algebraic Riccati equation (ARE) is the optimal solution \cite{luo2020policy}. The RL methods can be used to find the near-optimal solutions of the above optimality equations by stepwise iteration.

Currently, most RL algorithms are designed by two basic iteration frameworks, i.e., policy iteration (PI) and value iteration (VI) \cite{luo2020policy}. In \cite{jiang2012computational}, a model-free PI algorithm for CT linear systems is proposed to achieve optimal control. Gao {\it et al.} \cite{7444144} extends it to the model-free optimal output regulation (OOR) problem. The DT version of them is given by Jiang {\it et al.} \cite{jiang2019optimal}. Bian {\it et al.} \cite{bian2016value} proposed a model-free VI algorithm for CT linear systems, and Jiang {\it et al.} \cite{jiang2023reinforcement} extended their approach to the $H_{\infty}$ OOR problem for multiagent systems. Chen {\it et al.} \cite{chen2022robust} proposed a model-free robust OOR algorithm based on PI and input-output data for DT linear systems. The methods, including but not limited to the above, are classical model-free RL algorithms based on PI and VI. Moreover, the $\mathcal{Q}$-learning techniques are also developed based on VI and PI.

Recently, $\mathcal{Q}$-learning which is one of the classical techniques in RL has been developed to address optimal control problems \cite{al2007model,kiumarsi2014reinforcement,lee2012integral,vamvoudakis2017q,rizvi2018output,9640530}. $\mathcal{Q}$-learning algorithms are usually designed via an action-valued function that is defined according to the Bellman equation, often referred to as the $\mathcal{Q}$-function. PI and VI-based $\mathcal{Q}$-learning algorithms for DT linear systems were proposed by Kiumarsi {\it et al.} \cite{kiumarsi2014reinforcement} and Al-Tamimi {\it et al.} \cite{al2007model} respectively. And in \cite{lee2012integral,vamvoudakis2017q}, $\mathcal{Q}$-learning algorithms of CT linear systems have been developed. Subsequently, Rizvi {\it et al.} \cite{rizvi2018output} proposed a VI-based output feedback $\mathcal{Q}$-learning algorithm for DT linear systems. In \cite{rizvi2018output}, Rizvi {\it et al.} developed two output-feedback $\mathcal{Q}$-learning methods based on PI and VI for the linear quadratic regulator problem of DT systems. An {\it et al.} \cite{9640530} extended the $\mathcal{Q}$-learning approach to handle the distributed sensor scheduling problem with $H_{\infty}$ performance for wireless sensor networks. Recently, in \cite{lopez2023efficient}, Lopez {\it et al.} proposed an efficient off-policy $\mathcal{Q}$-learning and analyzed the performance of the algorithm.

Moreover, RL methods are also mainly categorized into on-policy and off-policy \cite{kiumarsi2017h,chen2022robust}. In the on-policy learning approach, the behavioral policy used to generate learning data is the same as the target policy used for decision making. The off-policy learning differs from it by differentiating between behavioral and target policies. Notably, $\mathcal{Q}$-learning is off-policy which uses the $\mathcal{Q}$-value of the next state and greedy action to update its $\mathcal{Q}$-values \cite{kiumarsi2017h}.
\subsection{Motivation}
Analyzing the above classical results, it is easy to conclude that PI and VI have different characteristics. The PI method must start with a known stabilizing control policy, but is characterized by fast convergence \cite{luo2020policy,chen2023adaptive}. Instead, the VI method begins with an easily determined value function with an arbitrary bounded control policy. But, VI method converges more slowly \cite{luo2020policy}. In each iteration, the PI algorithm obtains the real-valued function of the target control policy based on the known stabilizing control policy, which in turn updates the control policy, while the VI algorithm updates the control policy only after the value function has converged. This means that the target policy is obtained faster by the PI method with a known stabilizing control policy than the VI method. In \cite{luo2020policy,8122054}, the characteristics of PI and VI and their different convergence rates are analyzed in detail. It should be emphasized that obtaining stabilizing control is difficult. The process usually requires full a priori knowledge \cite{chen2023adaptive}. Although the method for seeking stabilizing control is provided in \cite{khargonekar1990robust}, it is still required that the nominal system be known. Moreover, It is important to develop efficient model-free off-policy techniques that do not require the current estimate of the optimal control policy to be applied to the system to re-collect data, which greatly improves efficiency.
Therefore, our aim is to study an efficient model-free off-policy algorithm and relax the initial control condition.

\subsection{Related Work}
Existing relevant results fall into two main categories, one that addresses the above problem by combining the advantages of VI and PI \cite{8122054,luo2019balancing,gao2022resilient,10388382}, and one that directly seeks a stabilizing control to satisfy the initial condition of PI \cite{chen2022homotopic,chen2023adaptive,10634984}. Luo {\it et al.} \cite{8122054} proposed multi-step heuristic dynamic programming to realize the trade-off between PI and VI. Subsequently, Luo {\it et al.} combined PI and VI by varying the coefficients  \cite{luo2019balancing}. Recently, a hybrid iteration method has been proposed by Gao {\it et al.} \cite{gao2022resilient}, which obtains a stabilizing control policy through VI, then converges quickly to the optimal solution through PI. Shen {\it et al.} \cite{10388382} extended this method to optimal control of unknown fast-sampling singularly perturbed systems. In \cite{jiang2022bias}, Jiang {\it et al.} developed a bias-PI algorithm for CT systems that can relax the initial stabilizing control condition by adding a bias parameter. In \cite{chen2022homotopic,chen2023adaptive}, Chen {\it et al.} proposed homotopy-based PI methods for linear and nonlinear CT systems, combining the homotopy method with PI to seek a stabilizing control for the unknown systems. In \cite{10634984}, Li {\it et al.} further extend the linear method to the $H_{\infty}$ cooperative OOR for CT multiagent systems. The method does not rely on VI to find the initial stabilizing control and converges faster. This homotopy-based approach originates from \cite{broussard1983active}, which gradually approaches a stable closed-loop system from a stable artificial system. The technology is further developed in \cite{feng2020connectivity,feng2020escaping,lamperski2020computing,chen2022homotopic,chen2023adaptive,10634984}. Although the method for calculating the stabilizing control of the CT system is given in \cite{feng2020connectivity,feng2020escaping}, a priori knowledge is used in them. The method for solving the stabilizing control of a DT linear system is given in \cite{lamperski2020computing} by using discount factor, but the updating rule of the discount factor and the iteration step size are not explicit. Unlike, the performance of the proposed homotopy-based model-free methods \cite{chen2022homotopic,chen2023adaptive}, which not only obtain the stabilizing control but also give an explicit iteration step size for the homotopy gain, are analyzed. Although the problem has been well solved for CT systems in \cite{chen2022homotopic,chen2023adaptive}, designing an optimal control method that can compute stabilizing control gains and is characterized by fast convergence remains a challenge for completely unknown DT linear systems.

\subsection{Contribution}
Inspired by the above results, this paper proposes two different PI-based model-free stabilizing optimal control algorithms for DT linear systems. Different from \cite{hewer1971iterative,chen2022robust,jiang2019optimal,kiumarsi2017h,fan2019model,kiumarsi2014reinforcement,lewis2010reinforcement,kiumarsi2015optimal,rizvi2018output}, the stabilizing control policy is not required as the initial condition in the two algorithms and they are characterized by fast convergence. First, a stable artificial system is designed by using damping coefficients, and it is gradually iterated to the original system by varying the damping coefficients. This allows the stabilizing control policy of the original closed-loop system to be obtained in a finite number of iteration steps. Then, two model-free versions of this method are designed by means of the off-policy iteration approach and off-policy $\mathcal{Q}$-learning, and the explicit selection criterions for the damping coefficients are given without using a priori knowledge of the model. The optimal control solution is obtained quickly after the stabilizing control gain is obtained. Note that in the two proposed algorithms, the process of calculating the stabilizing control gain is not only model-free, but the convergence boundary of the closed-loop spectral radius can be user-defined. In the proposed algorithms, the current estimates of the optimal control gain are not applied to the system to re-collect data, and the proposed algorithms are efficient.

\subsection{Organization}
Section \ref{section:2} describes the considered system and control objective and recalls existing model-based PI algorithm for DT systems. Section \ref{section:3} presents the model-based PI method for calculating a stabilizing control gain. Section \ref{section:4} proposes a stabilizing off-policy iteration algorithm and its convergence analysis. Section \ref{section:5} presents an off-policy $\mathcal{Q}$-learning algorithm and its convergence analysis. Section \ref{section:6} gives the simulation of an open-loop unstable system for two algorithms. Section \ref{section:7} summarizes the study results.

\subsection{Notation}
$\rho(\star)$ denotes the spectral radius of the matrix ``$\star$''. $\sigma_{\max}(\star)$ and $\sigma_{\min}(\star)$ denote the maximum and minimum singular values of the matrix ``$\star$''. For matrix $X\in\mathbb{R}^{n\times m}$, $vec(X)=[X_{1}^{T},X_{2}^{T},\ldots,X_{m}^{T}]^{T}\in\mathbb{R}^{nm}$, where $X_{i}\in\mathbb{R}^{n}$ is $i$th column of matrix $X$ for $i=1,\ldots,m$.
 If matrix $X=X^{T}\in\mathbb{R}^{n\times n}$, $vecs(X)=[X_{1,1},2X_{1,2},\ldots,2X_{1,n},X_{2,2},2X_{2,3},\ldots,2X_{n-1,n},X_{n,n}]^{T}\in\mathbb{R}^{\frac{1}{2}n(n+1)}$, where $X_{i,j}$ is the $i$th row and $j$th column element of matrix $X$.
For vector $\chi\in\mathbb{R}^{n}$, $vecv(\chi)=[\chi_{1}^{2},\chi_{1}\chi_{2},\ldots,\\
\chi_{1}\chi_{n},\chi_{2}^{2},\chi_{2}\chi_{3},\ldots,\chi_{n-1}\chi_{n},\chi_{n}^{2}]^{T}\in\mathbb{R}^{\frac{n(n+1)}{2}}$.
$\otimes$ denotes the Kronecker product. $I_{n}\in\mathbb{R}^{n\times n}$ denotes the identity matrix.

\section{Formulation and Preliminaries}\label{section:2}
\subsection{System and Problem Description}
Consider a DT linear system
\begin{equation} \label{1}
\begin{aligned}
&x(k+1)=Ax(k)+Bu(k)
\end{aligned}
\end{equation}
where $x\in\mathbb{R}^{n_x}$ is the state vector and $u\in\mathbb{R}^{n_u}$ is the input. $A\in\mathbb{R}^{n_x\times{n_x}}$ and $B\in\mathbb{R}^{n_x\times{n_u}}$ are unknown constant matrices.
\begin{assumption}
The pair $(A, B)$ is stabilizable.
\end{assumption}


The optimal control is to obtain a feedback controller
\begin{equation} \label{2}
\begin{aligned}
u=-Kx(k)
\end{aligned}
\end{equation}
such that system \eqref{1} is stabilized, where gain matrix $K\in\mathbb{R}^{n_u \times n_x}$ is obtained by minimizing the cost function
\begin{equation} \label{3a}
\begin{aligned}
J(x,u)=\sum_{i=0}^{\infty}\big(x(i)^{T}Qx(i)+u(i)^{T}Ru(i)\big),
\end{aligned}
\end{equation}
where $Q\in\mathbb{R}^{n_x\times{n_x}}>0$ and $R\in\mathbb{R}^{n_u\times{n_u}}>0$. Define the value function as
\begin{equation} \label{3}
\begin{aligned}
V(k)=\sum_{i=k}^{\infty}\big(x(i)^{T}Qx(i)+u(i)^{T}Ru(i)\big),
\end{aligned}
\end{equation}
that is, the cost after time $k$.
The problem is summarized as
\begin{equation} \label{3b}
\begin{aligned}
\quad\quad \min_{u}&\{\sum_{i=k}^{\infty}\big(x(i)^{T}Qx(i)+u(i)^{T}Ru(i)\big)\}\\
&s.t. \quad\eqref{1} \quad \& \quad \eqref{2}.
\end{aligned}
\end{equation}

According to Lyapunov theory \cite{dullerud2013course}, if the closed-loop system spectral radius satisfies $\rho(A-BK)<1$, then system \eqref{1} is Schur stable.

{\it Objective:}  Design a model-free optimal control method to compute the stabilizing control gain of system \eqref{1} and solve problem \eqref{3b}.
\subsection{Discrete-time Policy Iteration}
For system \eqref{1}, the value function \eqref{3} is denoted as the following quadratic closed-loop form
\begin{equation} \label{4}
\begin{aligned}
V(k)=x(k)^{T}Px(k),
\end{aligned}
\end{equation}
where $P=P^{T}>0$. According to \cite{lewis2012optimal}, the optimal feedback control for solving problem \eqref{3b} is given as
\begin{equation} \label{5}
\begin{aligned}
u=-(R+B^{T}PB)^{-1}B^{T}PAx(k)=-K^*x(k)
\end{aligned}
\end{equation}
where $P$ satisfies the ARE
\begin{equation} \label{6}
\begin{aligned}
A^{T}PA-P-A^{T}PB(R+B^{T}PB)^{-1}B^{T}PA=-Q
\end{aligned}
\end{equation}
To avoid solving DT ARE \eqref{6} directly, the classical model-based PI method is given in Lemma \ref{L1}.
\begin{lemma}\label{L1}\cite{hewer1971iterative,chen2022robust}
Given any initial stabilizing control gain $K^{0}$ satisfying $\rho(A-BK^{0})<1$. For $i=0,1,2,\dots$, solve $P^{i}=(P^{i})^{T}$ by the Lyapunov equation
\begin{equation} \label{7}
\begin{aligned}
(A-BK^{i})^{T}P^{i}(A-BK^{i})-P^{i}=-Q-(K^{i})^{T}RK^{i}.
\end{aligned}
\end{equation}
Update the gain by
\begin{equation} \label{8}
\begin{aligned}
K^{i+1}=(R+B^{T}P^{i}B)^{-1}B^{T}P^{i}A.
\end{aligned}
\end{equation}
Then, 1) $\rho(A-BK^{i+1})<1$; 2) $P^*\leq P^{i+1}\leq P^{i}$; 3) $\lim_{i\rightarrow\infty}P^{i}=P^*$, $\lim_{i\rightarrow\infty}K^{i}=K^*$.
\end{lemma}
\begin{remark}
For Lemma \ref{L1}, the initial stabilizing gain $K^{0}$ and the system $(A, B)$ are necessary to be known. Currently, finding a stabilizing control gain also relies on a priori knowledge of $(A, B)$. To realize model-free optimal control, it is critical to obtain an initial stabilizing control gain without using model information. If the initial stabilizing control is obtained, then the approximate optimal solution to the DT ARE \eqref{6} can be obtained by the traditional PI-based model-free algorithms such as \cite{jiang2019optimal,kiumarsi2014reinforcement}.
\end{remark}

\section{Model-based Policy Iteration for Stabilizing Control}\label{section:3}
First, we consider that $A$ and $B$ are available. A model-based method is designed to obtain a stabilizing control.

Define a positive constant $\bar{\rho}$ as
\begin{equation} \label{9}
\begin{aligned}
\bar{\rho}=\frac{1}{\rho(A)}.
\end{aligned}
\end{equation}
There exists a small constant $\beta$ satisfying $\bar{\rho}>\beta>0$ such that $\rho[(\bar{\rho}-\beta)(A-BK)]<1$ if $K=0$. Then, the system
\begin{equation} \label{10}
\begin{aligned}
x(k+1)=(\bar{\rho}-\beta)(Ax(k)+Bu(k))
\end{aligned}
\end{equation}
can be stabilized when $K=0$. To distinguish the process from Lemma \ref{L1}, define new matrices $\tilde{P}^{j}$ and $\tilde{K}^{j}$, where $j=0,1,2,\ldots$. Then, by Lemma \ref{L1}, if $\tilde{K}^{0}=0$, the matrix $\tilde{P}^{j}=(\tilde{P}^{j})^{T}>0$ is found from the Lyapunov function
\begin{equation} \label{11}
\begin{aligned}
&[(\bar{\rho}-\beta)(A-B\tilde{K}^{j})]^{T}\tilde{P}^{j}[(\bar{\rho}-\beta)(A-B\tilde{K}^{j})]-\tilde{P}^{j}\\
&=-Q-(\tilde{K}^{j})^{T}R\tilde{K}^{j}.
\end{aligned}
\end{equation}
Viewing $(\bar{\rho}-\beta)A$ and $(\bar{\rho}-\beta)B$ as $A$ and $B$ respectively in Lemma \ref{L1}, the control gain is given by $\tilde{K}^{j+1}=(\bar{\rho}-\beta)^{2}(R+(\bar{\rho}-\beta)^{2}B^{T}\tilde{P}^{j}B)^{-1}B^{T}\tilde{P}^{j}A$ for $j=0,1,2,\ldots$.

This process ensures that the artificial system $x(k+1)=(\bar{\rho}-\beta)(A-B\tilde{K}^{j})x(k)$ is stabilized, but it does not ensure that the original system \eqref{1} is stabilized. The following Theorem \ref{L2} provides a way to obtain stabilizing control gains by gradually iterating the artificial system \eqref{10} to the original closed-loop system.
\begin{theorem}\label{L2}
Initialize $\tilde{K}^{0}=0$, $R>0$, $Q>0$ and constants $\bar{\rho}>\beta>\alpha_{0}>0$. Find $\tilde{P}^{j}$ from
\begin{equation} \label{12}
\begin{aligned}
&[(\tilde{\beta}+\sum_{m=0}^{j}\alpha_{m})(A-B\tilde{K}^{j})]^{T}\tilde{P}^{j}[(\tilde{\beta}+\sum_{m=0}^{j}\alpha_{m})(A-B\tilde{K}^{j})]\\
&-\tilde{P}^{j}=-Q-(\tilde{K}^{j})^{T}R\tilde{K}^{j}
\end{aligned}
\end{equation}
where $\tilde{\beta}=\bar{\rho}-\beta$. Update $\tilde{K}^{j+1}$ and $\alpha_{j+1}$ by
\begin{equation} \label{13}
\begin{aligned}
\tilde{K}^{j+1}=&(\tilde{\beta}+\sum_{m=0}^{j}\alpha_{m})^{2}[R+(\tilde{\beta}+\sum_{m=0}^{j}\alpha_{m})^{2}B^{T}\tilde{P}^{j}B]^{-1}\\
&\times B^{T}\tilde{P}^{j}A,
\end{aligned}
\end{equation}
\begin{equation} \label{14}
\begin{aligned}
0<\alpha_{j+1}<\frac{1}{\rho(A-B\tilde{K}^{j+1})}-(\tilde{\beta}+\sum_{m=0}^{j}\alpha_{m}),
\end{aligned}
\end{equation}
for $j=0,1,2,\ldots$.
Then, there are
\begin{enumerate}
\item{For $j=1,2,\ldots$, $\rho(A-B\tilde{K}^{j})<1/(\tilde{\beta}+\sum_{m=0}^{j}\alpha_{m})$.}
\item{If $\alpha_{j}$ is bounded, then $\alpha_{j+1}$ is also bounded.}
\end{enumerate}
\end{theorem}

{\it Proof.}
First, we prove 1) by using induction. For $j=0$, since $\bar{\rho}>\beta>\alpha_{0}>0$ and $\tilde{K}^{0}=0$, there is $\rho[(\tilde{\beta}+\alpha_{0})(A-B\tilde{K}^{0})]<\bar{\rho}\rho(A)=1$. Viewing $(\tilde{\beta}+\alpha_{0})A$ and $(\tilde{\beta}+\alpha_{0})B$ as $A$ and $B$ in Lemma \ref{L1}, there is $\rho[(\tilde{\beta}+\alpha_{0})(A-B\tilde{K}^{1})]<1$. Then, there is $1/\rho(A-B\tilde{K}^{1})-(\tilde{\beta}+\alpha_{0})>0$. Therefore, there must exist $\alpha_{1}$ satisfying \eqref{14}. From \eqref{14}, one has $\rho[(\tilde{\beta}+\alpha_{0}+\alpha_{1})(A-B\tilde{K}^{1})]<1$. The statement 1) holds for $j=1$.

Assume that statement 1) holds for $j = a$. Then, there are $\rho[(\tilde{\beta}+\sum_{m=0}^{a}\alpha_{m})(A-B\tilde{K}^{a})]<1$ and a matrix $\tilde{P}^{a}=(\tilde{P}^{a})^{T}>0$ that is solved by
\begin{equation} \label{15}
\begin{aligned}
&[(\tilde{\beta}+\sum_{m=0}^{a}\alpha_{m})(A-B\tilde{K}^{a})]^{T}\tilde{P}^{a}[(\tilde{\beta}+\sum_{m=0}^{a}\alpha_{m})(A-B\tilde{K}^{a})]\\
&-\tilde{P}^{a}=-Q-(\tilde{K}^{a})^{T}R\tilde{K}^{a}.
\end{aligned}
\end{equation}
By using \eqref{13} and \eqref{15}, we can obtain the gain matrix as
\begin{equation}
\begin{aligned} \nonumber
&\tilde{K}^{a+1}\\
&=(\tilde{\beta}+\sum_{m=0}^{a}\alpha_{m})^{2}[R+(\tilde{\beta}+\sum_{m=0}^{a}\alpha_{m})^{2}B^{T}\tilde{P}^{a}B]^{-1}B^{T}\tilde{P}^{a}A.
\end{aligned}
\end{equation}
Clearly, $\tilde{K}^{a}$ is a stabilizing control of the system $(\tilde{\beta}+\sum_{m=0}^{a}\alpha_{m})(A-B\tilde{K}^{a})$, and if $(\tilde{\beta}+\sum_{m=0}^{a}\alpha_{m})A$ and $(\tilde{\beta}+\sum_{m=0}^{a}\alpha_{m})B$ are taken as $A$ and $B$ in Lemma \ref{L1}, one obtains
\begin{equation} \label{16}
\begin{aligned}
\rho[(\tilde{\beta}+\sum_{m=0}^{a}\alpha_{m})(A-B\tilde{K}^{a+1})]<1.
\end{aligned}
\end{equation}
From \eqref{16}, it follows that $1/(\rho(A-B\tilde{K}^{a+1}))-(\tilde{\beta}+\sum_{m=0}^{a}\alpha_{m})>0$. Then, there exists at least one positive constant $\alpha_{a+1}$ satisfying
\begin{equation} \label{17}
\begin{aligned}
0<\alpha_{a+1}<\frac{1}{\rho(A-B\tilde{K}^{a+1})}-(\tilde{\beta}+\sum_{m=0}^{a}\alpha_{m})
\end{aligned}
\end{equation}
such that $\rho[(\tilde{\beta}+\sum_{m=0}^{a+1}\alpha_{m})(A-B\tilde{K}^{a+1})]<1$. This also shows that there exists a positive definite solution to \eqref{12} for $k = a+1$. Therefore, statement 1) holds for $k=a+1$. In summary, statement 1) is established.

Proof of statement 2). Since $A$, $B$, $R$ and $Q$ are constant matrices and $\alpha_0$ is bounded, $\tilde{P}^{0}$ is also bounded by \eqref{12}. Further according to \eqref{13} one can get that $\tilde{K}^{1}$ is bounded. Recursively, if $\alpha_j$ is bounded, one obtains that $\tilde{K}^{j+1}$ is bounded. Then, $1/(\rho(A-B\tilde{K}^{a+1}))-(\tilde{\beta}+\sum_{m=0}^{a+1}\alpha_{m})>0$ is bounded. Finally, it can be obtained that $\alpha_{j+1}$ is bounded.$\Box$

\begin{remark}
It is worth noting that Theorem \ref{L2} degenerates to Lemma \ref{L1} if setting $\tilde{\beta}=1$, $\alpha_{j}=0$ and $\tilde{K}^{0}$ satisfies $\rho(A-B\tilde{K}^{0})<1$. In the statement 1) of Theorem \ref{L2}, $\rho(A-B\tilde{K}^{j})<1/(\tilde{\beta}+\sum_{m=0}^{j}\alpha_{m})$ is satisfied during the iterations. There exists a positive integer $\underline{j}$ such that $\tilde{\beta}+\sum_{m=0}^{j}\alpha_{m}\geq1$ when $j>\underline{j}$. Then, $\rho(A-B\tilde{K}^{j})<1$, i.e., $\tilde{K}^{j}$ is a stabilizing control gain if $j>\underline{j}$. Obviously, if a larger $\alpha_{j+1}$ is chosen to satisfy \eqref{14}, the iteration will be faster. Moreover, it can be obtained that $1/(\tilde{\beta}+\sum_{m=0}^{j}\alpha_{m})$ is a designed bound on the closed-loop spectral radius $\rho(A-B\tilde{K}^{j})$ from statement 1) of Theorem \ref{L2}.
\end{remark}

\begin{figure}[htbp]
      \centering
      \includegraphics[width=9cm,height=2.8cm]{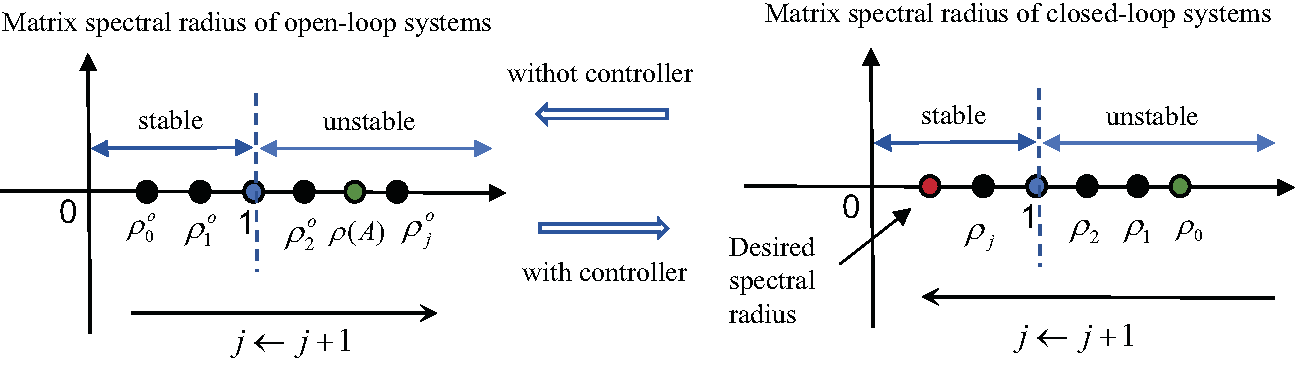}
      \caption{The spectral radius of the open-loop artificial system $(\tilde{\beta}+\sum_{m=0}^{j}\alpha_{m})A$ is denoted as $\rho^{o}_{j}=\rho[(\tilde{\beta}+\sum_{m=0}^{j}\alpha_{m})A]$; The spectral radius of the closed-loop system $A-B\tilde{K}^{j}$ is denoted as $\rho_{j}=\rho(A-B\tilde{K}^{j})$.}\label{fig1}
\end{figure}
\begin{remark}
Fig. \ref{fig1} illustrates the significance of the proposed method and Theorem \ref{L2}. As the process in Theorem \ref{L2} is iterated, the open-loop artificial system $(\tilde{\beta}+\sum_{m=0}^{j}\alpha_{m})A$ gradually approaches the open-loop original system $A$ and equals the original system when $\tilde{\beta}+\sum_{m=0}^{j}\alpha_{m}=1$, and the closed-loop spectral radius $\rho(A-B\tilde{K}^{j})$ decreases gradually. Until $\tilde{\beta}+\sum_{m=0}^{j}\alpha_{m}=1$, the spectral radius of the original closed-loop system $A-B\tilde{K}^{j}$ is iterated into the stable range.
\end{remark}
\section{Model-free Policy Iteration to Slove Discrete-time  ARE}\label{section:4}
For Theorem \ref{L2}, a prior knowledge of $(A, B)$ is required. Moreover, choosing the damping coefficients $\tilde{\beta}$ and $\alpha_{j+1}$ also depends on the known $A$ and $B$. Therefore, in this section, these model-based restrictions are removed.

\subsection{Collect Data and Determine $\tilde{\beta}$ Without A Priori Knowledge}\label{section:4:1}
To ensure stability, $\tilde{\beta}$ should satisfy $0<\tilde{\beta}<\bar{\rho}$, where $\tilde{\beta}=\bar{\rho}-\beta$ and $\beta>0$ is a small constant. For convenience, we can directly design the appropriate $\tilde{\beta}$.

Defining
\begin{equation} \label{18}
\begin{aligned}
\tilde{A}^{j}=(\tilde{\beta}+\sum_{m=0}^{j}\alpha_{m})(A-B\tilde{K}^{j}),
\end{aligned}
\end{equation}
then system \eqref{1} is written as
\begin{equation} \label{19}
\begin{aligned}
x(k+1)=(\tilde{\beta}+\sum_{m=0}^{j}\alpha_{m})^{-1}\tilde{A}^{j}x(k)+B(\tilde{K}^{j}x(k)+u(k)).
\end{aligned}
\end{equation}
By \eqref{19}, we have
\begin{equation} \label{20}
\begin{aligned}
&x(k+1)^{T}\tilde{P}^{j}x(k+1)-x(k)^{T}\tilde{P}^{j}x(k)\\
=&(\tilde{\beta}+\sum_{m=0}^{j}\alpha_{m})^{-2}x(k)^{T}\tilde{A}^{jT}\tilde{P}^{j}\tilde{A}^{j}x(k)+2(\tilde{\beta}+\sum_{m=0}^{j}\alpha_{m})^{-1}\\
&\times x(k)^{T}\tilde{A}^{jT}\tilde{P}^{j}B(\tilde{K}^{j}x(k)+u(k))-x(k)^{T}\tilde{P}^{j}x(k)\\
&+(\tilde{K}^{j}x(k)+u(k))^{T}B^{T}\tilde{P}^{j}B(\tilde{K}^{j}x(k)+u(k)).
\end{aligned}
\end{equation}
Substituting \eqref{12} and \eqref{18} into \eqref{20} gives
\begin{equation} \label{21}
\begin{aligned}
&x(k+1)^{T}\tilde{P}^{j}x(k+1)-x(k)^{T}\tilde{P}^{j}x(k)\\
=&(\tilde{\beta}+\sum_{m=0}^{j}\alpha_{m})^{-2}x(k)^{T}(-Q-(\tilde{K}^{j})^{T}R\tilde{K}^{j})x(k)\\
&+2x(k)^{T}\tilde{L}_{1}^{j}(\tilde{K}^{j}x(k)+u(k))-x(k)^{T}\tilde{K}^{jT}\tilde{L}_{2}^{j}\tilde{K}^{j}x(k)\\
&-[1-(\tilde{\beta}+\sum_{m=0}^{j}\alpha_{m})^{-2}]x(k)^{T}\tilde{P}^{j}x(k)
+u(k)^{T}\tilde{L}_{2}^{j}u(k),
\end{aligned}
\end{equation}
where
\begin{equation} \label{21a}
\begin{aligned}
\tilde{L}_{1}^{j}=A^{T}\tilde{P}^{j}B, \quad \tilde{L}_{2}^{j}=B^{T}\tilde{P}^{j}B.
\end{aligned}
\end{equation}
Collecting the data yields the following matrices as
\begin{equation} \label{22}
\begin{aligned}
\Xi_{1}=[&vecv(x(k_{0}+1))-vecv(x(k_{0})),\ldots,\\
&vecv(x(k_{s}+1))-vecv(x(k_{s}))]^{T},\\
\Xi_{2}=[&x(k_{0})\otimes x(k_{0}),x(k_{1})\otimes x(k_{1}),\ldots,x(k_{s})\otimes x(k_{s})]^{T},\\
\Xi_{3}=[&u(k_{0})\otimes x(k_{0}),u(k_{1})\otimes x(k_{1}),\ldots,u(k_{s})\otimes x(k_{s})]^{T},\\
\Xi_{4}=[&vecv(u(k_{0})),vecv(u(k_{1})),\ldots,vecv(u(k_{s}))]^{T},\\
\Xi_{5}=[&vecv(x(k_{0})),vecv(x(k_{1})),\ldots,vecv(x(k_{s}))]^{T},\\
\Xi_{6}=[&vecv(\tilde{K}^{j}x(k_{0})),\ldots,vecv(\tilde{K}^{j}x(k_{s}))]^{T},
\end{aligned}
\end{equation}
where $k_{0}<k_{1}<\ldots<k_{s}$.
Using \eqref{21}-\eqref{22}, one has
\begin{equation} \label{25}
\begin{aligned}
\psi^{j}[vecs(\tilde{P}^{j})^{T},vec(\tilde{L}_{1}^{j})^{T},vecs(\tilde{L}_{2}^{j})^{T}]^{T}=\phi^{j},
\end{aligned}
\end{equation}
where $\psi^{j}=\big[\Xi_{1}-((\tilde{\beta}+\sum_{m=0}^{j}\alpha_{m})^{-2}-1)\Xi_{5}, -2\Xi_{2}(I_{n_x}\otimes \tilde{K}^{jT})-2\Xi_{3},-\Xi_{4}+\Xi_{6}\big]$ and $\phi^{j}=(\tilde{\beta}+\sum_{m=0}^{j}\alpha_{m})^{-2}\Xi_{5}vecs(-Q-(\tilde{K}^{j})^{T}R\tilde{K}^{j})$.

\begin{lemma}\label{rank1}
Collecting data such that
\begin{equation} \label{26}
\begin{aligned}
rank([\Xi_{2},\Xi_{3},\Xi_{4}])= \frac{1}{2}(n_{x}+n_{u})(n_{x}+n_{u}+1),
\end{aligned}
\end{equation}
then $\psi^{j}$ has full column-rank.
\end{lemma}

{\it Proof.}
See Appendix \ref{app1}.$\Box$

By Lemma \ref{rank1} and the least squares, if \eqref{26} is satisfied, the unique solution to \eqref{25} is obtained as
\begin{equation} \label{27}
\begin{aligned}
&[vecs(\tilde{P}^{j})^{T},vec(\tilde{L}_{1}^{j})^{T},vecs(\tilde{L}_{2}^{j})^{T}]^{T}=(\psi^{jT}\psi^{j})^{-1}\psi^{jT}\phi^{j}.
\end{aligned}
\end{equation}


To determine $\tilde{\beta}$, consider the case $j = 0$. Then, we have $\tilde{K}^{0}=0$. When $j = 0$, the data is collected to satisfy \eqref{26}. By using \eqref{25}, we can find
\begin{equation} \label{28}
\begin{aligned}
&[vecs(\tilde{P}^{0})^{T},vec(\tilde{L}_{1}^{0})^{T},vecs(\tilde{L}_{2}^{0})^{T}]^{T}=(\psi^{0T}\psi^{0})^{-1}\psi^{0T}\phi^{0},
\end{aligned}
\end{equation}
where $\psi^{0}=\big[\Xi_{1}-((\tilde{\beta}+\alpha_{0})^{-2}-1)\Xi_{5}, -2\Xi_{2}(I_{n_x}\otimes \tilde{K}^{0T})-2\Xi_{3},-\Xi_{4}+\Xi_{6}\big]$ and $\phi^{0}=(\tilde{\beta}+\alpha_{0})^{-2}\Xi_{5}vecs(-Q-(\tilde{K}^{0})^{T}R\tilde{K}^{0})$.

Recalling Section \ref{section:3}, the designed $\tilde{\beta}$ and $\alpha_0$ should satisfy the model-based condition
\begin{equation} \label{29}
\begin{aligned}
\tilde{\beta}+\alpha_0<\bar{\rho}=\frac{1}{\rho(A)}.
\end{aligned}
\end{equation}
Since $\bar{\rho}$ depends on a priori knowledge of $A$, it cannot be derived directly. Therefore, we should design a model-free condition equivalent to \eqref{29}. According to the linear systems theory, if the solution $\tilde{P}^{j}$ to the Lyapunov function \eqref{12} of the artificial system $(\tilde{\beta}+\sum_{m=0}^{j}\alpha_{m})(A-B\tilde{K}^{j})$ is positive definite, it follows that the artificial system $(\tilde{\beta}+\sum_{m=0}^{j}\alpha_{m})(A-B\tilde{K}^{j})$ is Schur, i.e., $\rho[(\tilde{\beta}+\sum_{m=0}^{j}\alpha_{m})(A-B\tilde{K}^{j})]<1$. Therefore, when $\tilde{P}^{0}$ is positive from \eqref{28}, it can be deduced that $\tilde{\beta}+\alpha_{0}$ satisfies $\rho[(\tilde{\beta}+\alpha_{0})(A)]<1$.

{\it Determine $\tilde{\beta}$ by the model-free method:}
Set $\tilde{\beta}\leftarrow\tilde{\beta}_{z}$ and sufficiently small $\alpha_{0}>0$, where $\tilde{\beta}_{z}$ is a decreasing constant and satisfies
\begin{equation} \label{31}
\begin{aligned}
0<\tilde{\beta}_{z+1}<\tilde{\beta}_{z},\quad \lim_{z\rightarrow+\infty}\tilde{\beta}_{z}=0, 
\end{aligned}
\end{equation}
for $z=0,1,2,\ldots.$. For step $z$, if $\tilde{P}^{0}>0$ from \eqref{28}, then $\tilde{\beta}_{z}$ is used as $\tilde{\beta}$ in \eqref{27} and \eqref{28}, otherwise repeat \eqref{31} for $z\leftarrow z+1$ until $\tilde{P}^{0}>0$. The final output $\tilde{\beta}\leftarrow\tilde{\beta}_{z}$.

\begin{remark}
Through the above process, $\tilde{\beta}$ and $\alpha_{0}$ are determined such that $\tilde{\beta}+\alpha_{0}<1/\rho(A)$, where $\alpha_{0}$ is set directly to a sufficiently small positive constant.
$\tilde{\beta}$ is a damping coefficient that is usually set $\tilde{\beta}\in(0,1)$ for open-loop unstable systems with $\tilde{K}^{0}=0$. When the $\tilde{\beta}$ is set to satisfy $\tilde{\beta}\in(0,\bar{\rho}-\alpha_{0})$, then a larger $\tilde{\beta}$ accelerates the convergence of the spectral radius $\rho(A-B\tilde{K}^{j})$. When the $\tilde{\beta}$ is set to satisfy $\tilde{\beta}\in(\bar{\rho}-\alpha_{0},1)$, then the above model-free process (i.e., Phase 1 of Algorithms \ref{alg1} and \ref{alg2}) quickly adjusts the $\tilde{\beta}$ to the interval $(0,\bar{\rho}-\alpha_{0})$.
Moreover, choosing $\alpha_{j+1}$ according to \eqref{14} ensures that $\tilde{K}^{j+1}$ is a stabilizing control gain of system $(\tilde{\beta}+\sum_{m=0}^{j+1}\alpha_{m})(A-B\tilde{K}^{j+1})$ for step $j+1$, and \eqref{14} is subsequently replaced with the model-free version.
\end{remark}
\subsection{Calculate Stabilizing Gain $\tilde{K}^{j+1}$ and Select Damping Coefficient $\alpha_{j+1}$}
From \eqref{14}, the choice of $\alpha_{j+1}$ depends on known $(A,B)$. Therefore, this subsection proposes an equivalent selection criterion of $\alpha_{j+1}$ without a priori knowledge.

{\it Update gain $\tilde{K}^{j+1}$ by the model-free method:}
According to \eqref{27}, $L_{1}^{j}$ and $L_{2}^{j}$ are obtained. Then, from \eqref{13}, we can update the control gain matrix by
\begin{equation} \label{32}
\begin{aligned}
\tilde{K}^{j+1}=(\tilde{\beta}+\sum_{m=0}^{j}\alpha_{m})^{2}\big(R+(\tilde{\beta}+\sum_{m=0}^{j}\alpha_{m})^{2}\tilde{L}_{2}^{j}\big)^{-1}\tilde{L}_{1}^{jT}.
\end{aligned}
\end{equation}

{\it Select $\alpha_{j+1}$ by the model-free method:}
The following selection criterion of $\alpha_{j+1}$ is designed to replace \eqref{14} in Theorem \ref{L2} as
\begin{equation} \label{33}
\begin{aligned}
0<&\alpha_{j+1}\\
<&(\tilde{\beta}+\sum_{m=0}^{j}\alpha_{m})\sqrt{\frac{\sigma_{\min}[Q+(\tilde{K}^{j+1})^{T}R\tilde{K}^{j+1}]}{\sigma_{\max}[\tilde{P}^{j}-Q-(\tilde{K}^{j+1})^{T}R\tilde{K}^{j+1}]}+1}\\
&-(\tilde{\beta}+\sum_{m=0}^{j}\alpha_{m}).
\end{aligned}
\end{equation}

Unlike \eqref{14}, the system model matrices $(A,B)$ are not involved in \eqref{33}. For step $j+1$, $\tilde{P}^{j}$ and $\tilde{K}^{j+1}$ have been obtained. $\alpha_{j+1}$ satisfies \eqref{33} such that $\tilde{K}^{j+1}$ is a stabilizing control gain of $(\tilde{\beta}+\sum_{m=0}^{j+1}\alpha_{m})(A-B\tilde{K}^{j+1})$. The analysis is given below.

\begin{theorem}\label{the2}
Using control gain $\tilde{K}^{j+1}$ from \eqref{13} and $\alpha_{j+1}$ from \eqref{33}, there exists the unique positive solution $\tilde{P}^{j+1}$ to \eqref{12} for step $j+1$, and $\tilde{K}^{j+1}$ is also a stabilizing control gain of closed-loop artificial system $(\tilde{\beta}+\sum_{m=0}^{j+1}\alpha_{m})(A-B\tilde{K}^{j+1})$.
\end{theorem}

{\it Proof.}
If $(\tilde{\beta}+\sum_{m=0}^{j}\alpha_{m})A$ and $(\tilde{\beta}+\sum_{m=0}^{j}\alpha_{m})B$ are taken as $A$ and $B$ in Lemma \ref{L1}, one obtains
$\rho[(\tilde{\beta}+\sum_{m=0}^{j}\alpha_{m})(A-B\tilde{K}^{j+1})]<1$. Then, $\tilde{K}^{j+1}$ is a stabilizing control gain of $(\tilde{\beta}+\sum_{m=0}^{j}\alpha_{m})(A-B\tilde{K}^{j+1})$. Further, by Lemma \ref{L1}, there must exist a positive-definite matrix $\hat{\tilde{P}}^{j+1}=(\hat{\tilde{P}}^{j+1})^{T}$ can be found from
\begin{equation} \label{34}
\begin{aligned}
&[(\tilde{\beta}+\sum_{m=0}^{j}\alpha_{m})(A-B\tilde{K}^{j+1})]^{T}\hat{\tilde{P}}^{j+1}[(\tilde{\beta}+\sum_{m=0}^{j}\alpha_{m})(A\\
&-B\tilde{K}^{j+1})]-\hat{\tilde{P}}^{j+1}=-Q-(\tilde{K}^{j+1})^{T}R\tilde{K}^{j+1}
\end{aligned}
\end{equation}
and $0<\hat{\tilde{P}}^{j+1}\leq{\tilde{P}}^{j}$. This gives
\begin{equation} \label{35}
\begin{aligned}
&(A-B\tilde{K}^{j+1})^{T}\hat{\tilde{P}}^{j+1}(A-B\tilde{K}^{j+1})\\
&=(\tilde{\beta}+\sum_{m=0}^{j}\alpha_{m})^{-2}[\hat{\tilde{P}}^{j+1}-Q-(\tilde{K}^{j+1})^{T}R\tilde{K}^{j+1}].
\end{aligned}
\end{equation}
From \eqref{18}, one has $\tilde{A}^{j+1}=(\tilde{\beta}+\sum_{m=0}^{j+1}\alpha_{m})(A-B\tilde{K}^{j+1})$. Then, it is easy to get
\begin{equation} \label{36}
\begin{aligned}
&(\tilde{A}^{j+1})^{T}\hat{\tilde{P}}^{j+1}\tilde{A}^{j+1}-\hat{\tilde{P}}^{j+1}\\
\overset{(a)}{=}&(\frac{\tilde{\beta}+\sum_{m=0}^{j+1}\alpha_{m}}{\tilde{\beta}+\sum_{m=0}^{j}\alpha_{m}})^{2}[\hat{\tilde{P}}^{j+1}-Q-(\tilde{K}^{j+1})^{T}R\tilde{K}^{j+1}]-\hat{\tilde{P}}^{j+1}\\
\overset{(b)}{\leq}&\big((1+\frac{\alpha_{j+1}}{\tilde{\beta}+\sum_{m=0}^{j}\alpha_{m}})^{2}-1\big)[{\tilde{P}}^{j}-Q-(\tilde{K}^{j+1})^{T}R\tilde{K}^{j+1}]\\
&-Q-(\tilde{K}^{j+1})^{T}R\tilde{K}^{j+1}\\
\overset{(c)}{<}&\sigma_{\min}[Q+(\tilde{K}^{j+1})^{T}R\tilde{K}^{j+1}]I_{n_{x}}-Q-(\tilde{K}^{j+1})^{T}R\tilde{K}^{j+1}\\
<&0,
\end{aligned}
\end{equation}
where $(a)$ is obtained by the definition of $\tilde{A}^{j+1}$ and \eqref{35}, $(b)$ is obtained by $0<\hat{\tilde{P}}^{j+1}\leq{\tilde{P}}^{j}$ and $(c)$ is obtained by \eqref{33}. Therefore, there is $\rho(\tilde{A}^{j+1})<1$ and $\tilde{K}^{j+1}$ is a stabilizing control gain of system $\tilde{A}^{j+1}$. Then, there exists the unique positive solution $\tilde{P}^{j+1}$ to \eqref{12} for step $j+1$. Since $\rho(\tilde{A}^{j+1})<1$, one has
$\rho(A-B\tilde{K}^{j+1})<1/(\tilde{\beta}+\sum_{m=0}^{j+1}\alpha_{m})$. This also yields \eqref{14}. As can be seen from the above process, \eqref{14} can be easily obtained by using the selection criteria \eqref{33} and after relaxation. Therefore, the range \eqref{33} is contained in \eqref{14}.
$\Box$

\begin{remark}\label{rem5}
Explicit selection criterion of damping coefficient $\alpha_{j+1}$ is established without a priori knowledge. Then, a stabilizing control gain $\tilde{K}^{j+1}$ of system \eqref{1} can be obtained. Moreover, Theorem \ref{the2} and its proof show that choosing the damping coefficient $\alpha_{j+1}$ according to \eqref{33} is equivalent to \eqref{14}. If a larger $\alpha_{j+1}$ is chosen, a stabilizing control gain $\tilde{K}^{j+1}$ is obtained faster and $\rho(A-B\tilde{K}^{j+1})$ converges faster. However, this does not mean that choosing a larger $\alpha_{j+1}$ will result in a smaller spectral radius, because the iteration stops immediately when $\tilde{\beta}+\sum_{m=0}^{j+1}\alpha_{m}\geq1$ is reached.
\end{remark}

\subsection{The Overall Stabilizing Off-Policy Iteration Algorithm}
Suppose that stabilizing control gain $\tilde{K}^{j+1}$ has been obtained by the previous method.
Using ${K}^{0}=\tilde{K}^{j+1}$ and ${K}^{i}$, we write system \eqref{1} as
\begin{equation} \label{a}
\begin{aligned}
x(k+1)=A^{i}x(k)+B({K}^{i}x(k)+u(k)),
\end{aligned}
\end{equation}
where $A^{i}=(A-B{K}^{i})$.
By \eqref{a}, we have
\begin{equation} \label{b}
\begin{aligned}
&x(k+1)^{T}{P}^{i}x(k+1)-x(k)^{T}{P}^{i}x(k)\\
=&x(k)^{T}{A}^{iT}{P}^{i}{A}^{i}x(k)+2x(k)^{T}{A}^{T}{P}^{}B({K}^{i}x(k)+u(k))\\
&+({K}^{i}x(k)+u(k))^{T}B^{T}P^{i}B({K}^{i}x(k)+u(k))\\
&-x(k)^{T}{P}^{i}x(k)
\end{aligned}
\end{equation}
Substituting \eqref{7} and \eqref{a} into \eqref{b} gives
\begin{equation} \label{c}
\begin{aligned}
&x(k+1)^{T}{P}^{i}x(k+1)-x(k)^{T}{P}^{i}x(k)\\
=&x(k)^{T}(-Q-({K}^{i})^{T}R{K}^{i})x(k)\\
&+2x(k)^{T}A^{T}{P}^{i}B({K}^{i}x(k)+u(k))\\
&-x(k)^{T}{K}^{iT}B^{T}{P}^{i}B{K}^{i}x(k)+u(k)^{T}B^{T}{P}^{i}Bu(k)
\end{aligned}
\end{equation}
Define
\begin{equation} \label{d}
\begin{aligned}
{L}_{1}^{i}=A^{T}{P}^{i}B, \quad {L}_{2}^{i}=B^{T}{P}^{i}B.
\end{aligned}
\end{equation}
Using the data matrices $\Xi_{1}$, $\Xi_{2}$, $\Xi_{3}$, $\Xi_{4}$ and $\Xi_{5}$ collected in \eqref{22} and $\Xi_{6}=[vecv({K}^{i}x(k_{0})),\ldots,vecv({K}^{i}x(k_{s}))]^{T}$, the matrices $\Psi^{i}=\big[\Xi_{1}, -2\Xi_{2}(I_{n_x}\otimes {K}^{iT})-2\Xi_{3},-\Xi_{4}+\Xi_{6}\big]$ and $\Phi^{i}=\Xi_{5}vecs(-Q-({K}^{i})^{T}R{K}^{i})$ are obtained.
Then, \eqref{c} is rewritten as
\begin{equation} \label{f}
\begin{aligned}
\Psi^{i}[vecs({P}^{i})^{T},vec({L}_{1}^{i})^{T},vecs({L}_{2}^{i})^{T}]^{T}=\Phi^{i}.
\end{aligned}
\end{equation}

\begin{lemma}\label{rank2}
If \eqref{26} is satisfied, $\Psi^{i}$ has full column-rank.
\end{lemma}
{\it Proof.}
It is similar to Lemma \ref{rank1}.
$\Box$

By Lemma \ref{rank2}, then \eqref{f} has a unique solution as
\begin{equation} \label{g}
\begin{aligned}
&[vecs({P}^{i})^{T},vec({L}_{1}^{i})^{T},vecs({L}_{2}^{i})^{T}]^{T}=(\Psi^{iT}\Psi^{i})^{-1}\Psi^{iT}\Phi^{i}.
\end{aligned}
\end{equation}
According to \eqref{8}, we have
\begin{equation} \label{h}
\begin{aligned}
{K}^{i+1}=(R+{L}_{2}^{i})^{-1}{L}_{1}^{iT}.
\end{aligned}
\end{equation}
Finally, the overall Algorithm \ref{alg1} and its convergence analysis are given as follows.
\begin{algorithm}[!h]
	\caption{Stabilizing Off-Policy Iteration Algorithm}\label{alg1}
	\begin{algorithmic}[!h]
		\STATE {\bf{Initialize:}} Set $j\leftarrow0$, $z\leftarrow0$, $\tilde{K}^{0}\leftarrow0$, two sufficiently small positive constants $\varepsilon_{1}$ and $\alpha_{0}$, the monotonically decreasing sequence $\{\tilde{\beta}_{z}\}_{z=0}^{\infty}\in(0,1)$ and bounded probing noise. Collect data until \eqref{26} is satisfied.
     \REPEAT
         \STATE Calculate $\tilde{P}^{0}$ from \eqref{28}.
         \STATE Select $\tilde{\beta}\leftarrow\tilde{\beta}_{z}$ by \eqref{31}, $z\leftarrow z+1$.
     \UNTIL {$\tilde{P}^{0}>0$}. ({\bf Phase 1 end})
     \REPEAT
         \STATE Solve $\tilde{P}^{j}$, $\tilde{L}_{1}^{j}$ and $\tilde{L}_{2}^{j}$ by \eqref{27}.
         \STATE Calculate $\tilde{K}^{j+1}$ by \eqref{32}.
         \STATE Choose $\alpha_{j+1}$ from \eqref{33}.
         \STATE {$j\leftarrow j+1$.}
     \UNTIL {$\tilde{\beta}+\sum_{m=0}^{j}\alpha_{m}\geq1$.} ({\bf Phase 2 end})
         \STATE {Set $K^{i}\leftarrow\tilde{K}^{j+1}$, $i\leftarrow0$.} \label{step12}
     \REPEAT
         \STATE Solve ${P}^{i}$, ${L}_{1}^{i}$ and ${L}_{2}^{i}$ from \eqref{g}.
         \STATE Calculate ${K}^{i+1}$ by \eqref{h}.
         \STATE $i\leftarrow i+1$.
     \UNTIL {$\| P^{i}- P^{i-1} \|< \varepsilon_{1}$.} ({\bf Phase 3 end})
         \STATE {\bf{Return:}} ${K}^{i+1}$ and ${P}^{i}$ as solutions to ARE \eqref{6}.
	\end{algorithmic}
\end{algorithm}
\begin{theorem}\label{the3}
Set $\tilde{K}^{0} = 0$, update $\tilde{P}^{j}$ and $\tilde{K}^{j+1}$ via \eqref{27} and \eqref{32} until $\tilde{\beta}+\sum_{m=0}^{j+1}\alpha_{m}\geq1$, respectively. The obtained $\tilde{K}^{j+1}$ is a stabilizing control gain of system \eqref{1} and the closed-loop system satisfies $\rho(A-B\tilde{K}^{j+1})<1/(\tilde{\beta}+\sum_{m=0}^{j+1}\alpha_{m})$ for $j=0,1,2,\ldots$. Setting $K^{0}=\tilde{K}^{j+1}$ and updating ${P}^{i}$ and ${K}^{i+1}$ by \eqref{g} and \eqref{h} for $i=0,1,\ldots$, respectively, there exist $\lim_{i\rightarrow\infty}P^{i}=P^*$ and $\lim_{i\rightarrow\infty}K^{i}=K^*$.
\end{theorem}

{\it Proof.}
If \eqref{26} is satisfied, \eqref{25} is equivalent to \eqref{21}. The unique $\tilde{P}^{j}$, $\tilde{L}_{1}^{j}$ and $\tilde{L}_{2}^{j}$ can be solved from \eqref{25} and the unique $\tilde{K}^{j+1}$ can be obtained according to \eqref{32}. From \eqref{32}, it follows that $\tilde{K}^{j+1}$ satisfies \eqref{13}. Theorem \ref{the2} shows that selecting $\alpha_{j+1}$ via \eqref{33} is equivalent to selecting $\alpha_{j+1}$ via \eqref{14}. Then, it is equivalent to solving $\tilde{P}^{j}$ and $\tilde{K}^{j+1}$ from Theorem \ref{L2}. Thus, the closed-loop system satisfies Property 1) in Theorem \ref{L2}. If $\tilde{\beta}+\sum_{m=0}^{j+1}\alpha_{m}\geq1$, the obtained $\tilde{K}^{j+1}$ is a stabilizing control gain of system \eqref{1}.

Set $K^{0}=\tilde{K}^{j+1}$ as the initial stabilizing gain. Phase 3 of Algorithm \ref{alg1} is an off-policy algorithm based on Lemma \ref{L1}. If condition \eqref{26} is satisfied, \eqref{f} is equivalent to \eqref{c}. Then, solving ${P}^{i}$ from \eqref{f} is equivalent to solving ${P}^{i}$ from \eqref{7}. According to \eqref{h}, it is obtained that $K^{i}$ satisfies \eqref{8}. This is equivalent to finding ${P}^{i}$ and $K^{i}$ by Lemma \ref{L1}. Therefore, there exist $\lim_{i\rightarrow\infty}P^{i}=P^*$ and $\lim_{i\rightarrow\infty}K^{i}=K^*$. 
$\Box$

\begin{remark}
In Algorithm \ref{alg1}, the data need only be collected once until the rank condition \eqref{26} is satisfied. After Phases 1 and 2, the stabilizing gain $\tilde{K}^{j+1}$ is obtained adaptively. The solutions to ARE \eqref{6} are obtained adaptively by Phase 3. The obtained estimated value ${K}^{i}$ (or $\tilde{K}^{j}$) is not applied to the system to re-collect data.
\end{remark}

\section{Off-Policy $\mathcal{Q}$-Learning Algorithm Design}\label{section:5}
To further extend Theorem \ref{L2}, an efficient off-policy $\mathcal{Q}$-learning algorithm is developed in this section.

\subsection{Select $\tilde{\beta}$ and $\alpha_{j+1}$ and Calculate Stabilizing Gain $\tilde{K}^{j+1}$ by $\mathcal{Q}$-Learning}\label{section:5:1}
According to system \eqref{1}, the following artificial system is obtained as
\begin{equation} \label{37}
\begin{aligned}
\bar{x}(k+1)=(\tilde{\beta}+\sum_{m=0}^{j}\alpha_{m})A{x}(k)+(\tilde{\beta}+\sum_{m=0}^{j}\alpha_{m})Bu(k)
\end{aligned}
\end{equation}
where $\bar{x}(k+1)=(\tilde{\beta}+\sum_{m=0}^{j}\alpha_{m}){x}(k+1)$.
For system \eqref{37}, we consider the value functions
\begin{equation} \label{38}
\begin{aligned}
\tilde{V}(x(k))=&{x}(k)^{T}\tilde{P}{x}(k),\\
\tilde{V}(\bar{x}(k+1))=&\bar{x}(k+1)^{T}\tilde{P}\bar{x}(k+1),
\end{aligned}
\end{equation}
 where $\tilde{P}=\tilde{P}^{T}>0$. The Bellman equation for system \eqref{37} is obtained as
\begin{equation} \label{39}
\begin{aligned}
\tilde{V}(x(k))
=&{x}(k)^{T}Q{x}(k)+u(k)^{T}Ru(k)+\tilde{V}(\bar{x}(k+1)).
\end{aligned}
\end{equation}

Define the discrete-time $\mathcal{Q}$-function as $\tilde{\mathcal{Q}}(x(k),u(k))=\tilde{V}(x(k))$. According to \eqref{39}, one has
\begin{equation} \label{40}
\begin{aligned}
\tilde{\mathcal{Q}}(x(k),u(k))=&{x}(k)^{T}Q{x}(k)+u(k)^{T}Ru(k)\\
&+\tilde{V}(\bar{x}(k+1)).
\end{aligned}
\end{equation}
For convenience, let $\gamma_{j}=\tilde{\beta}+\sum_{m=0}^{j}\alpha_{m}$.
Substituting \eqref{37}, \eqref{38} and \eqref{39} to \eqref{40}, there is
\begin{equation} \label{41}
\begin{aligned}
\tilde{\mathcal{Q}}(x(k),u(k))=&{x}(k)^{T}Q{x}(k)+u(k)^{T}Ru(k)\\
&+\gamma_{j}^{2}(Ax(k)+Bu(k))^{T}\tilde{P}(Ax(k)+Bu(k))\\
=&X(k)^{T}\tilde{H}X(k)\\
=&X(k)^{T}\left[
\begin{array}{cc}
\tilde{H}_{xx}& \tilde{H}_{xu}\\
\tilde{H}_{ux}& \tilde{H}_{uu}\\
\end{array}
\right]X(k)
\end{aligned}
\end{equation}
where $X(k)=[x(k)^{T},u(k)^{T}]^{T}$, $\tilde{H}=\tilde{H}^{T}$ and
\begin{equation} \label{42}
\begin{aligned}
&\tilde{H}_{xx}=\gamma_{j}^{2}A^{T}\tilde{P}A+Q,\quad\tilde{H}_{uu}=\gamma_{j}^{2}B^{T}\tilde{P}B+R.\\
&\tilde{H}_{xu}=\gamma_{j}^{2}A^{T}\tilde{P}B, \quad \tilde{H}_{ux}=\tilde{H}_{xu}^{T}\\
\end{aligned}
\end{equation}
Using \eqref{41}, \eqref{42} and $\partial\tilde{\mathcal{Q}}(x(k),u(k))/\partial u_{k}=0$, we have
\begin{equation} \label{42b1}
\begin{aligned}
u(k)=-(\tilde{H}_{uu})^{-1}\tilde{H}_{ux}x(k).
\end{aligned}
\end{equation}
Then, it is obvious to get
\begin{equation} \label{42b}
\begin{aligned}
\tilde{K}=(\tilde{H}_{uu})^{-1}\tilde{H}_{ux}.
\end{aligned}
\end{equation}
Using \eqref{40}, one obtains
\begin{equation} \label{42d}
\begin{aligned}
X(k)^{T}\tilde{H}X(k)
&=X(k)^{T}\left[
\begin{array}{cc}
Q& 0\\
0& R\\
\end{array}
\right]X(k)\\
&+\gamma_{j}^{2}\left[
\begin{array}{c}
x(k+1)\\
-\tilde{K}x(k+1)\\
\end{array}
\right]^{T}\tilde{H}\left[
\begin{array}{c}
x(k+1)\\
-\tilde{K}x(k+1)\\
\end{array}
\right].
\end{aligned}
\end{equation}
Letting $\mathcal{M}=\gamma_{j}\left[\begin{array}{c}
I_{n_x}\\
-\tilde{K}\\
\end{array}
\right][A\quad B]$ and using $x(k+1)=Ax(k)+Bu(k)$, it can be obtained that \eqref{42d} is equivalent to the Lyapunov equation
\begin{equation} \label{42c}
\begin{aligned}
\tilde{H}
=\left[
\begin{array}{cc}
Q& 0\\
0& R\\
\end{array}
\right]+\mathcal{M}^{T}\tilde{H}\mathcal{M}
\end{aligned}
\end{equation}
\begin{lemma}\label{lem4}
Lyapunov equation \eqref{42c} has a unique positive definite solution $\tilde{H}$ if and only if $\gamma_{j}(A-B\tilde{K})$ is stable.
\end{lemma}

{\it Proof.}
See Appendix \ref{app2}.
$\Box$

According to \eqref{42b} and \eqref{42d}, the following PI-based $\mathcal{Q}$-learning process can be designed.

{\it 1). Policy evaluation:} Given the appropriate $\tilde{\beta}$ and $\alpha_{0}$. For $j=0,1,2,\ldots$, the matrix $\tilde{H}^{j}$ of the $\mathcal{Q}$-function can be solved by
\begin{equation} \label{43}
\begin{aligned}
&X(k)^{T}\tilde{H}^{j}X(k)\\
=&X(k)^{T}\left[
\begin{array}{cc}
Q& 0\\
0& R\\
\end{array}
\right]X(k)+(\tilde{\beta}+\sum_{m=0}^{j}\alpha_{m})^{2}\\
&\times\left[\begin{array}{c}
x(k+1)\\
-\tilde{K}^{j}x(k+1)\\
\end{array}
\right]^{T}\tilde{H}^{j}\left[\begin{array}{c}
x(k+1)\\
-\tilde{K}^{j}x(k+1)\\
\end{array}
\right]
\end{aligned}
\end{equation}

{\it 2). Policy improvement:} The control gain can be updated by
\begin{equation} \label{43a}
\begin{aligned}
\tilde{K}^{j+1}=(\tilde{H}_{uu}^{j})^{-1}\tilde{H}_{ux}^{j}.
\end{aligned}
\end{equation}

{\it 3). Select $\alpha_{j+1}$:}
Update $\alpha_{j+1}$ by \eqref{33}, where $\tilde{P}^{j}$ is given by
\begin{equation} \label{44}
\begin{aligned}
\tilde{P}^{j}=\left[\begin{array}{c}
I_{n_x}\\
-\tilde{K}^{j}\\
\end{array}
\right]^{T}\tilde{H}^{j}\left[\begin{array}{c}
I_{n_x}\\
-\tilde{K}^{j}\\
\end{array}
\right].
\end{aligned}
\end{equation}

Next, use least squares to solve \eqref{43}.
The following matrices are obtained by collecting data as
\begin{equation} \label{46}
\begin{aligned}
&\Gamma_{X}=[vecv(X(k_{0})),vecv(X(k_{1})),\ldots,vecv(X(k_{s}))]^{T},\\
&\Gamma_{x}=[vecv(z(k_{0}+1)), vecv(z(k_{1}+1)),\ldots,
vecv(z(k_{s}+1))]^{T},\\
&\tilde{\Theta}^{j}=\Gamma_{X}-(\tilde{\beta}+\sum_{m=0}^{j}\alpha_{m})^{2}\Gamma_{x},\quad\xi^{j}=\Gamma_{X}vecs(\bar{Q}),\\
&\bar{Q}=\left[
\begin{array}{cc}
Q& 0\\
0& R\\
\end{array}
\right],
z(k+1)=[x(k+1)^{T},(\tilde{K}^{j}x(k+1))^{T}]^{T},
\end{aligned}
\end{equation}
where $k_{0}<k_{1}<\ldots<k_{s}$.
From \eqref{43}, one has
\begin{equation} \label{47}
\begin{aligned}
\tilde{\Theta}^{j}vecs(\tilde{H}^{j})=\xi^{j}.
\end{aligned}
\end{equation}
Similar to Lemma \ref{rank1}, it can be deduced from \eqref{46} and \eqref{47} that if the collected system data satisfy
\begin{equation} \label{48}
\begin{aligned}
rank(\Gamma_{X})=\frac{1}{2}(n_{x}+n_{u})(n_{x}+n_{u}+1),
\end{aligned}
\end{equation}
matrix $\tilde{\Theta}^{j}$ has full column-rank. The unique solution to \eqref{47} is derived as
\begin{equation} \label{49}
\begin{aligned}
vecs(\tilde{H}^{j})=(\tilde{\Theta}^{jT}\tilde{\Theta}^{j})^{-1}\tilde{\Theta}^{jT}\xi^{j}.
\end{aligned}
\end{equation}

{\it Determine $\tilde{\beta}$:}
It follows from Lemma \ref{lem4} that if the derived $\tilde{H}^{0}$ is positive definite, then $(\tilde{\beta}+\alpha_{0})(A-B\tilde{K}^{0})$ is stable.
Updating $\tilde{\beta}$ still is based on \eqref{31}. If $j=0$, then $\tilde{\Theta}^{0}=\Gamma_{X}-(\tilde{\beta}+\alpha_{0})^{2}\Gamma_{x}$ and \eqref{47} becomes
\begin{equation} \label{50}
\begin{aligned}
\tilde{\Theta}^{0}vecs(\tilde{H}^{0})=\xi^{0}.
\end{aligned}
\end{equation}
If \eqref{48} is satisfied at $j = 0$, then the unique solution to \eqref{50} can be found as
\begin{equation} \label{51}
\begin{aligned}
vecs(\tilde{H}^{0})=(\tilde{\Theta}^{0T}\tilde{\Theta}^{0})^{-1}\tilde{\Theta}^{0T}\xi^{0}.
\end{aligned}
\end{equation}
If $\tilde{H}^{0}>0$ is satisfied, then $\tilde{\beta}$ is determined, otherwise, update $\tilde{\beta}$ by \eqref{31} until $\tilde{H}^{0}>0$.

After $\tilde{\beta}$ is determined, update $\tilde{H}^{j}$ via \eqref{49}, $\tilde{K}^{j+1}$ via \eqref{43a} and $\alpha_{j+1}$ via \eqref{44} and \eqref{33} until $\tilde{\beta}+\sum_{m=0}^{j+1}\alpha_{m}\geq1$. It has been analyzed in Section \ref{section:4:1} that the choice of $\alpha_{j+1}$ according to \eqref{33} ensures convergence. Since iterating the Lyapunov equation \eqref{43} is equivalent to iterating \eqref{12} (See Lemma \ref{L4}), the selection criteria \eqref{33} still guarantees the convergence of the method in this section.
\subsection{The Overall Stabilizing Off-Policy $\mathcal{Q}$-Learning Algorithm}
Suppose that stabilizing control gain $\tilde{K}^{j+1}$ has been obtained by \eqref{43a}. Set $K^{0}=\tilde{K}^{j+1}$. 

Consider the quadratic form value function of system \eqref{1} as $V(x(k))=x(k)^{T}Px(k)$ and
\begin{equation} \label{53a}
\begin{aligned}
V(x(k+1))=x(k+1)^{T}Px(k+1).
\end{aligned}
\end{equation}
Define the discrete-time $\mathcal{Q}$-function as $\mathcal{Q}({x}(k),{u}(k))=V(x(k))$. Then, we have
\begin{equation} \label{53}
\begin{aligned}
\mathcal{Q}({x}(k),{u}(k))=&{x}(k)^{T}Q{x}(k)+u(k)^{T}Ru(k)\\
&+V(x(k+1)).
\end{aligned}
\end{equation}
By using \eqref{1}, \eqref{53a} and \eqref{53},
it is easy to get
\begin{equation} \label{54}
\begin{aligned}
{\mathcal{Q}}(x(k),u(k))=&{x}(k)^{T}Q{x}(k)+u(k)^{T}Ru(k)\\
&+(Ax(k)+Bu(k))^{T}P(Ax(k)+Bu(k))\\
=&X(k)^{T}{H}X(k)\\
=&X(k)^{T}\left[
\begin{array}{cc}
{H}_{xx}& {H}_{xu}\\
{H}_{ux}& {H}_{uu}\\
\end{array}
\right]X(k)
\end{aligned}
\end{equation}
where $X(k)=[x(k)^{T},u(k)^{T}]^{T}$, ${H}={H}^{T}$ and
\begin{equation} \label{55}
\begin{aligned}
&{H}_{xx}=A^{T}{P}A+Q,\quad {H}_{xu}=A^{T}{P}B, \\
&\quad {H}_{ux}={H}_{xu}^{T},\quad {H}_{uu}=B^{T}{P}B+R.\\
\end{aligned}
\end{equation}
Let $\partial {\mathcal{Q}}(x(k),u(k))/\partial u(k)=0$ and $u(k)=-Kx(k)$, there is
\begin{equation} \label{55a}
\begin{aligned}
{K}=({H}_{uu})^{-1}{H}_{ux}.
\end{aligned}
\end{equation}
Similar to \eqref{42c}-\eqref{43a}, the following PI-based $\mathcal{Q}$-learning process is designed.

{\it 1). Policy evaluation:} Set $K^{0}=\tilde{K}^{j+1}$. For $i=0,1,2,\ldots$, the matrix ${H}^{i}$ of the $\mathcal{Q}$-function is updated by
\begin{equation} \label{56}
\begin{aligned}
&X(k)^{T}{H}^{i}X(k)\\
=&X(k)^{T}\left[
\begin{array}{cc}
Q& 0\\
0& R\\
\end{array}
\right]X(k)\\
&+\left[\begin{array}{c}
x(k+1)\\
-{K}^{i}x(k+1)\\
\end{array}
\right]^{T}{H}^{i}\left[\begin{array}{c}
x(k+1)\\
-{K}^{i}x(k+1)\\
\end{array}
\right]
\end{aligned}
\end{equation}

{\it 2). Policy improvement:} The control policy gain is updated by
\begin{equation} \label{57}
\begin{aligned}
{K}^{i+1}=({H}_{uu}^{i})^{-1}{H}_{ux}^{i}.
\end{aligned}
\end{equation}

Use the data matrices $\Gamma_{X}$ and $\Gamma_{x}$ collected in \eqref{46}, where $\tilde{K}^{j}$ is replaced by ${K}^{i}$ in $\Gamma_{x}$. Then,
the following matrices can be obtained as
\begin{equation} \label{58}
\begin{aligned}
\xi^{i}=\Gamma_{X}vecs(\bar{Q}),\quad {\Theta}^{i}=\Gamma_{X}-\Gamma_{x}.
\end{aligned}
\end{equation}
According to \eqref{56}, there is
\begin{equation} \label{59}
\begin{aligned}
{\Theta}^{i}vecs({H}^{i})=\xi^{i}
\end{aligned}
\end{equation}
If condition \eqref{48} is satisfied,
\eqref{59} can be uniquely solved as
\begin{equation} \label{61}
\begin{aligned}
vecs(H^{i})=({\Theta}^{iT}{\Theta}^{i})^{-1}{\Theta}^{iT}\xi^{i}.
\end{aligned}
\end{equation}
Further, update $K^{i+1}$ by \eqref{57}.
Finally, the overall Algorithm \ref{alg2} is given as follows.
\begin{algorithm}[!h]
	\caption{Stabilizing Off-Policy $\mathcal{Q}$-Learning Algorithm}\label{alg2}
	\begin{algorithmic}[1]
		\STATE {\bf{Initialize:}} Set $j\leftarrow0$, $z\leftarrow0$, $\tilde{K}^{0}\leftarrow0$, two sufficiently small positive constants $\varepsilon_{1}$ and $\alpha_{0}$, the monotonically decreasing sequence $\{\tilde{\beta}_{z}\}_{z=0}^{\infty}\in(0,1)$ and bounded probing noise. Collect data until \eqref{48} is satisfied.
      \REPEAT
         \STATE Solve $\tilde{H}^{0}$ by \eqref{51}.
	     \STATE Select $\tilde{\beta}=\tilde{\beta}_{z}$ by \eqref{31}.
      \UNTIL {$\tilde{H}^{0}>0$.}  ({\bf Phase 1 end})
      \REPEAT
         \STATE Solve $\tilde{H}^{j}$ by \eqref{49}.
         \STATE Calculate $\tilde{P}^{j}$ from \eqref{44}.
         \STATE Calculate $\tilde{K}^{j+1}$ from \eqref{43a}.
         \STATE Choose $\alpha_{j+1}$ from \eqref{33}.
         \STATE {$j\leftarrow j+1$.}
       \UNTIL $\tilde{\beta}+\sum_{m=0}^{j}\alpha_{m}\geq1$. ({\bf Phase 2 end})
         \STATE Set $K^{i}\leftarrow\tilde{K}^{j+1}$, $i\leftarrow0$.
       \REPEAT
         \STATE Solve ${H}^{i}$ by \eqref{61}.
         \STATE Calculate ${K}^{i+1}$ from \eqref{57}.
         \STATE {$i\leftarrow i+1$.}
       \UNTIL {$\| H^{i}- H^{i-1} \|< \varepsilon_{2}$.} ({\bf Phase 3 end})
         \STATE {\bf{Return:}} ${K}^{i+1}$ and $H^{i}$ as the optimal solutions.
	\end{algorithmic}
\end{algorithm}

The convergence is analyzed below. First, the following necessary Lemmas are given.
\begin{lemma}\label{lem2}
Solving \eqref{47} is equivalent to solving
\begin{equation} \label{62}
\begin{aligned}
\tilde{H}^{j}=&\bar{Q}+(\tilde{\beta}+\sum_{m=0}^{j}\alpha_{m})^{2}\\
&\times\left[\begin{array}{cc}
A&B\\
-\tilde{K}^{j}A & -\tilde{K}^{j}B\\
\end{array}
\right]^{T}\tilde{H}^{j}\left[\begin{array}{cc}
A&B\\
-\tilde{K}^{j}A & -\tilde{K}^{j}B\\
\end{array}
\right],
\end{aligned}
\end{equation}
and solving \eqref{59} is equivalent to solving
\begin{equation} \label{63}
\begin{aligned}
{H}^{i}=\bar{Q}+
\left[\begin{array}{cc}
A&B\\
-{K}^{i}A & -{K}^{i}B\\
\end{array}
\right]^{T}{H}^{i}\left[\begin{array}{cc}
A&B\\
-{K}^{i}A & -{K}^{i}B\\
\end{array}
\right].
\end{aligned}
\end{equation}
\end{lemma}

{\it Proof.}
If \eqref{48} is satisfied, then $\tilde{\Theta}^{j}$ has full column-rank. There is a one-to-one linear mapping relationship between \eqref{47} and \eqref{43}. Solving \eqref{47} is equivalent to solving \eqref{43}. Then, \eqref{47} is equivalent to
\begin{equation} \label{64}
\begin{aligned}
\tilde{H}^{j}
=\bar{Q}+\mathcal{M}_{j,\tilde{K}^{j}}^{T}\tilde{H}^{j}\mathcal{M}_{j,\tilde{K}^{j}}
\end{aligned}
\end{equation}
where $\mathcal{M}_{j,\tilde{K}^{j}}=(\tilde{\beta}+\sum_{m=0}^{j}\alpha_{m})\left[\begin{array}{c}
I_{n_x}\\
-\tilde{K}^{j}\\
\end{array}
\right][A\quad B]$. Thus, Solving for $\tilde{H}^{j}$ from \eqref{43} is equivalent to solving for $\tilde{H}^{j}$ from \eqref{62}. Similarly, it can be obtained that \eqref{59} is equivalent to \eqref{63}.
$\Box$

\begin{lemma}\label{L3}
Using \eqref{43} and \eqref{43a}, the matrices $\tilde{H}^{j}$ and $\tilde{K}^{j+1}$ can be written as
\begin{equation} \label{65}
\begin{aligned}
\tilde{H}^{j}=\left[
\begin{array}{cc}
\gamma_{j}^{2}A^{T}\tilde{P}^{j}A+Q& \gamma_{j}^{2}A^{T}\tilde{P}^{j}B\\
\gamma_{j}^{2}B^{T}\tilde{P}^{j}A& \gamma_{j}^{2}B^{T}\tilde{P}^{j}B+R\\
\end{array}
\right]
\end{aligned}
\end{equation}
and \eqref{13}, where $\tilde{P}^{j}$ is given by \eqref{44} and
$\tilde{\beta}+\sum_{m=0}^{j}\alpha_{m}$ is denoted as $\gamma_{j}$ for step $j$. Using \eqref{56} and \eqref{57}, the matrices ${H}^{i}$ and ${K}^{i+1}$ can be written as
\begin{equation} \label{66}
\begin{aligned}
{H}^{i}=\left[
\begin{array}{cc}
A^{T}{P}^{i}A+Q& A^{T}{P}^{i}B\\
B^{T}{P}^{i}A& B^{T}{P}^{i}B+R\\
\end{array}
\right]
\end{aligned}
\end{equation}
and \eqref{8}, where $P^{i}=[
I_{n_x}^{T},-{K}^{iT}]{H}^{i}[I_{n_x}^{T}, -{K}^{iT}]^{T}$.
\end{lemma}

{\it Proof.}
By \eqref{64}, it can be obtained that \eqref{65} holds. Similarly, \eqref{66} holds. Using \eqref{43a} and \eqref{57}, one gets \eqref{13} and \eqref{8}, respectively.
$\Box$

\begin{lemma}\label{L4}
Solving $\tilde{H}^{j}$ by \eqref{65} equals solving $\tilde{P}^{j}$ by
\begin{equation} \label{67}
\begin{aligned}
\tilde{P}^{j}=&\bar{A}^{jT}\tilde{P}^{j}\bar{A}^{j}+Q\\
&-\bar{A}^{jT}\tilde{P}^{j}\bar{B}^{j}(R+\bar{B}^{jT}\tilde{P}^{j}\bar{B}^{j})^{-1}\bar{B}^{jT}\tilde{P}^{j}\bar{A}^{j},
\end{aligned}
\end{equation}
where $\bar{A}^{j}=(\tilde{\beta}+\sum_{m=0}^{j}\alpha_{m})A$ and $\bar{B}^{j}=(\tilde{\beta}+\sum_{m=0}^{j}\alpha_{m})B$, and iterating ${H}^{i}$ by \eqref{66} is equivalent to iterating $P^{i}$ as
\begin{equation} \label{68}
\begin{aligned}
{P}^{i}=A^{T}P^{i}A+Q-A^{T}P^{i}B(R+B^{T}P^{i}B)^{-1}B^{T}P^{i}A.
\end{aligned}
\end{equation}
\end{lemma}

{\it Proof.}
There are $\tilde{P}^{j}=[I_{n_x}^{T},-\tilde{K}^{jT}]\tilde{H}^{j}[I_{n_x}^{T}, -\tilde{K}^{jT}]^{T}$ and
\begin{equation} \label{68a}
\begin{aligned}
P^{i}=\left[\begin{array}{c}
I_{n_x}\\
-{K}^{i}\\
\end{array}
\right]^{T}\tilde{H}^{j}\left[\begin{array}{c}
I_{n_x}\\
-{K}^{i}\\
\end{array}
\right].
\end{aligned}
\end{equation}
Substituting \eqref{65} into \eqref{44} gives
\begin{equation} \label{69}
\begin{aligned}
\tilde{P}^{j}=\gamma_{j}^{2}(A-B\tilde{K}^{j})^{T}\tilde{P}^{j}(A-B\tilde{K}^{j})+Q+(\tilde{K}^{j})^{T}R\tilde{K}^{j}.
\end{aligned}
\end{equation}
Using \eqref{13}, one has \eqref{67}. Similarly, substituting \eqref{66} into \eqref{68a} gives \eqref{7}. Then using \eqref{8}, one obtains \eqref{68}.
$\Box$

\begin{theorem}\label{the4}
Set $\tilde{K}^{0} = 0$ and update $\tilde{H}^{j}$, $\tilde{K}^{j+1}$ and $\alpha_{j+1}$ by \eqref{49}, \eqref{43a} and \eqref{33} until $\tilde{\beta}+\sum_{m=0}^{j+1}\alpha_{m}\geq1$ for $j=0,1,\ldots$, respectively. The obtained $\tilde{K}^{j+1}$ is a stabilizing control gain of system \eqref{1} and the closed-loop system satisfies $\rho(A-B\tilde{K}^{j+1})<1/(\tilde{\beta}+\sum_{m=0}^{j+1}\alpha_{m})$ for $j=0,1,2,\ldots$.  Setting $K^{0}=\tilde{K}^{j+1}$ and updating ${H}^{i}$ and ${K}^{i+1}$ by \eqref{61} and \eqref{57} for $i=0,1,2,\ldots$, respectively, gives $\lim_{i\rightarrow\infty}H^{i}=H^*$, where $H^*$ is solution to \eqref{54}. Then, ARE \eqref{6} is solved by ${K}^{i+1}$ and ${P}^{i}$.
\end{theorem}

{\it Proof.}
According to Lemmas \ref{lem2}-\ref{L4}, updating $\tilde{H}^{j}$ and $\tilde{K}^{j+1}$ by \eqref{49} and \eqref{43a} equals updating $\tilde{P}^{j}$ and $\tilde{K}^{j+1}$ via \eqref{12} and \eqref{14}. By Theorem \ref{the2}, choosing $\alpha_{j+1}$ according to \eqref{33} is equivalent to choosing $\alpha_{j+1}$ according to \eqref{14}. Therefore, the process is equivalent to Theorem \ref{L2}. According to Theorem \ref{L2}, the obtained $\tilde{K}^{j+1}$ is a stabilizing control gain of system \eqref{1} and the closed-loop spectral radius satisfies $\rho(A-B\tilde{K}^{j+1})<1/(\tilde{\beta}+\sum_{m=0}^{j+1}\alpha_{m})$ for $j=0,1,2,\ldots$.

Let $K^{0}=\tilde{K}^{j+1}$, Lemma \ref{L4} shows that updating $H^{i}$ and $K^{i+1}$ via \eqref{61} and \eqref{57} is equivalent to updating $P^{i}$ and $K^{i+1}$ via Lemma \ref{L1}. It is easy to obtain $\lim_{i\rightarrow\infty}H^{i}=H^*$ from \eqref{54} and Lemma \ref{L1}. Using \eqref{57} and \eqref{68a}, the $K^{i+1}$ and $P^{i}$ obtained are solutions to ARE \eqref{6}.
$\Box$
\begin{remark}
Unlike the traditional $\mathcal{Q}$-learning method \cite{al2007model,kiumarsi2014reinforcement,jiang2017tracking,rizvi2018output},
the input $u^{j}=-\tilde{K}^{j}x(k)$ (or $u^{i}=-{K}^{i}x(k)$) is never applied to the system. The proposed off-policy $\mathcal{Q}$-learning Algorithm \ref{alg2} requires only collecting system data once. Then, the data is used to find the stabilizing gain $\tilde{K}^{j+1}$ and the approximate optimal solution to $\mathcal{Q}$-function \eqref{54} respectively.
\end{remark}
\begin{remark}
The full column-rank conditions \eqref{26} and \eqref{48} are the conditions of satisfying persistence of excitation. The stabilizing control gains and optimal control gain can be online obtained by collecting data that satisfy the above conditions, while avoiding the system identification process as in \cite{pillonetto2025deep}. In addition, the policy iterations (Phase 3 of Algorithms \ref{alg1} and \ref{alg2}) have explicit termination conditions $\| P^{i}- P^{i-1} \|< \varepsilon_{1}$ and $\| H^{i}- H^{i-1} \|< \varepsilon_{2}$, and they guarantee the optimality of the resulting solutions. The damping coefficient-based policy iteration (Phases 1 and 2) also has an explicit termination condition $\tilde{\beta}+\sum_{m=0}^{j+1}\alpha_{m}\geq1$, which guarantees that the obtained control policy is stabilizing. Moreover, $1/(\tilde{\beta}+\sum_{m=0}^{j+1}\alpha_{m})$ serves as a convergence boundary for finding the stabilizing control gain, which can be set by the user.
\end{remark}
\begin{remark}
Similar to existing model-free RL methods (See References), the proposed algorithms do not require the system to be known, but are affected by the size of the system, such as the number of states and inputs. According to conditions \eqref{26} and \eqref{48}, it can be seen that if the number of states and inputs is larger, more input-state data need to be collected. The similar problem also exists in the system identification methods \cite{pillonetto2025deep}. Therefore, before using the proposed algorithm, it is necessary to first determine the number of states and inputs, which in turn establishes condition \eqref{26} (or condition \eqref{48}) and runs Algorithm \ref{alg1} (or Algorithm \ref{alg2}).
\end{remark}
\begin{remark}
Algorithm \ref{alg1} and Algorithm \ref{alg2} can be extended to two data-driven algorithms with specified convergence speeds. Design the constant $\delta>1$ as the rate of convergence and change the termination threshold for Phase 2 in Algorithms \ref{alg1} and \ref{alg2} to $\tilde{\beta}+\sum_{m=0}^{j+1}\alpha_{m}\geq\delta$. By Theorem \ref{L2}, the obtained $\tilde{K}^{j+1}$ satisfies $\rho(A-B\tilde{K}^{j+1})< \delta^{-1}$. Then, for $n=j+1,j+2,\ldots$, let $\tilde{\beta}+\sum_{m=0}^{n+1}\alpha_{m}=\delta$ and continue using the process in Phase 2 until $\|\tilde{P}^{n}- \tilde{P}^{n-1}\|<\epsilon_{1}$ or $\|\tilde{H}^{n}- \tilde{H}^{n-1}\|<\epsilon_{2}$. Then, the obtained $\tilde{K}^{n+1}$ and $\tilde{P}^{n}$ (or $\tilde{H}^{n}$) converge to the optimal solutions of the system $(\delta{A}, \delta{B})$ and $\tilde{K}^{n+1}$ is a stabilizing gain of system $({A}, {B})$ making $\rho(A-B\tilde{K}^{n+1})< \delta^{-1}$. Finally, the states of the system converge faster than $\delta^{-1}$. Note that they are not optimal control algorithms for the original system $({A}, {B})$. Due to space constraints, the two algorithms with specified convergence speeds and their analysis are not discussed in detail in this paper.
\end{remark}

\section{Simulation}\label{section:6}

Consider an open-loop unstable DT linear system as
\begin{equation} \label{70}
\begin{aligned}
x(k+1)=\left[
\begin{array}{cc}
-1 & 0.5\\
1.5 & 1.2\\
\end{array}
\right]x(k)+\left[
\begin{array}{c}
2\\
1.6\\
\end{array}
\right]u(k),
\end{aligned}
\end{equation}
where $x(k)=[x_{1}(k),x_{2}(k)]^{T}$.
Its open-loop poles are $-1.3$ and $1.5$. Set $Q=6I_{2}$ and $R=1$. Using $(A,B,Q,R)$ directly, the solution to ARE \eqref{6} and optimal gain are calculated as
\begin{equation} \label{71}
\begin{aligned}
P^*=\left[
\begin{array}{cc}
   27.8194  &  7.5337\\
    7.5337  &  8.7798
\end{array}
\right],&\ K^*=\left[
\begin{array}{cc}
     -0.1313  &  0.3759
\end{array}
\right].
\end{aligned}
\end{equation}
Further, the solution of the $\mathcal{Q}$-function \eqref{54} is calculated by \eqref{56} as
\begin{equation} \label{72}
\begin{aligned}
H^*=\left[
\begin{array}{ccc}
   30.9728 &  -1.4963  &-24.0202\\
   -1.4963  & 34.6382  & 68.7845\\
  -24.0202 &  68.7845 & 182.9699
\end{array}
\right].
\end{aligned}
\end{equation}
They can be used to verify the optimality of the results of Algorithms \ref{alg1} and \ref{alg2}. The initial states of the system are set as $x_{1}(0)=5$ and $x_{2}(0)=-5$.

\subsection{Validation of Model-Free PI Algorithm \ref{alg1}}\label{sec.6.1}
We set $\tilde{\beta}=0.1$, $\alpha_{0}=0.0001$ and choose the initial control input matrix as $\tilde{K}^{0}=[0,0]$. For $j=0,1,2,\ldots$, $\alpha_{j+1}$ is chosen as
\begin{equation} \label{72a}
\begin{aligned}
\alpha_{j+1}=a\times\bar{\alpha}
\end{aligned}
\end{equation}
where $a\in(0,1)$ is a coefficient and $\bar{\alpha}$ is the upper bound in \eqref{33}.

\begin{figure}
    \centering
	  \subfloat[]{
       \includegraphics[width=0.48\linewidth]{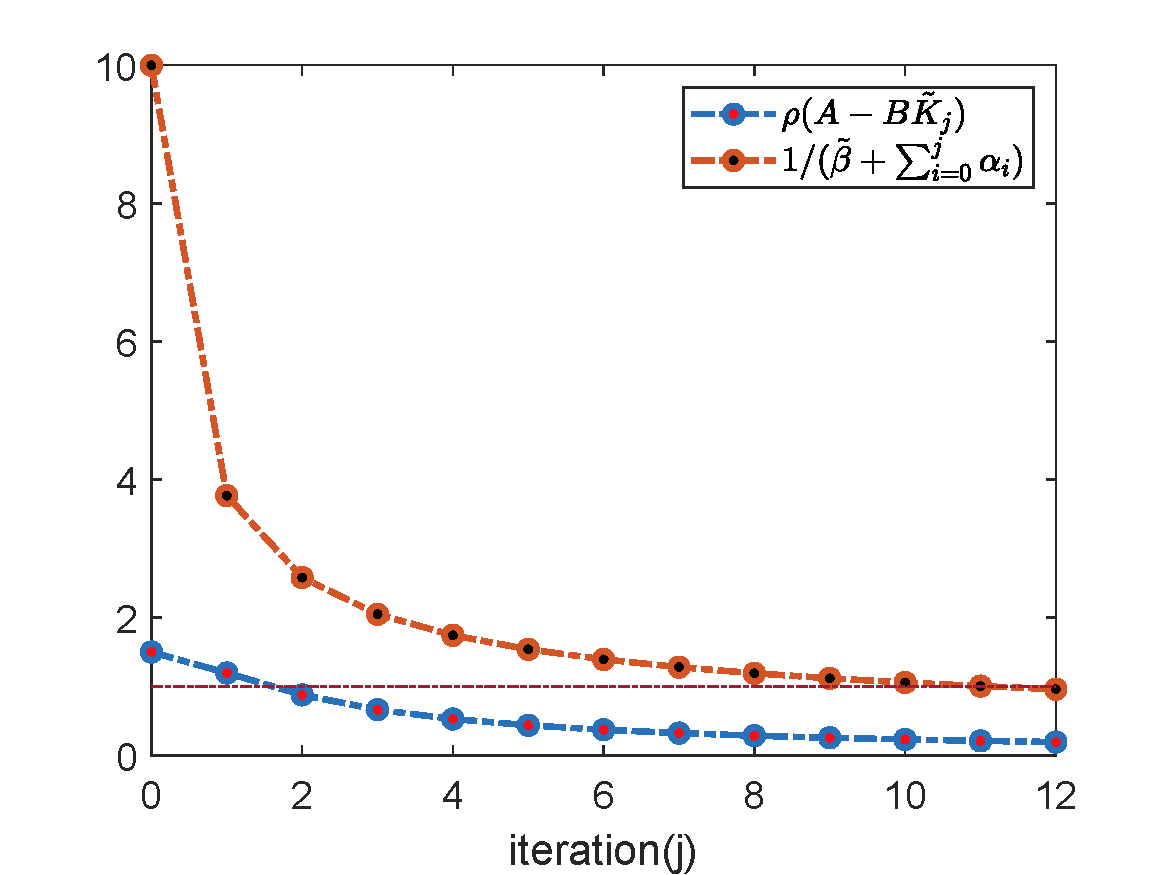}}
       \hfill
	  \subfloat[]{
        \includegraphics[width=0.48\linewidth]{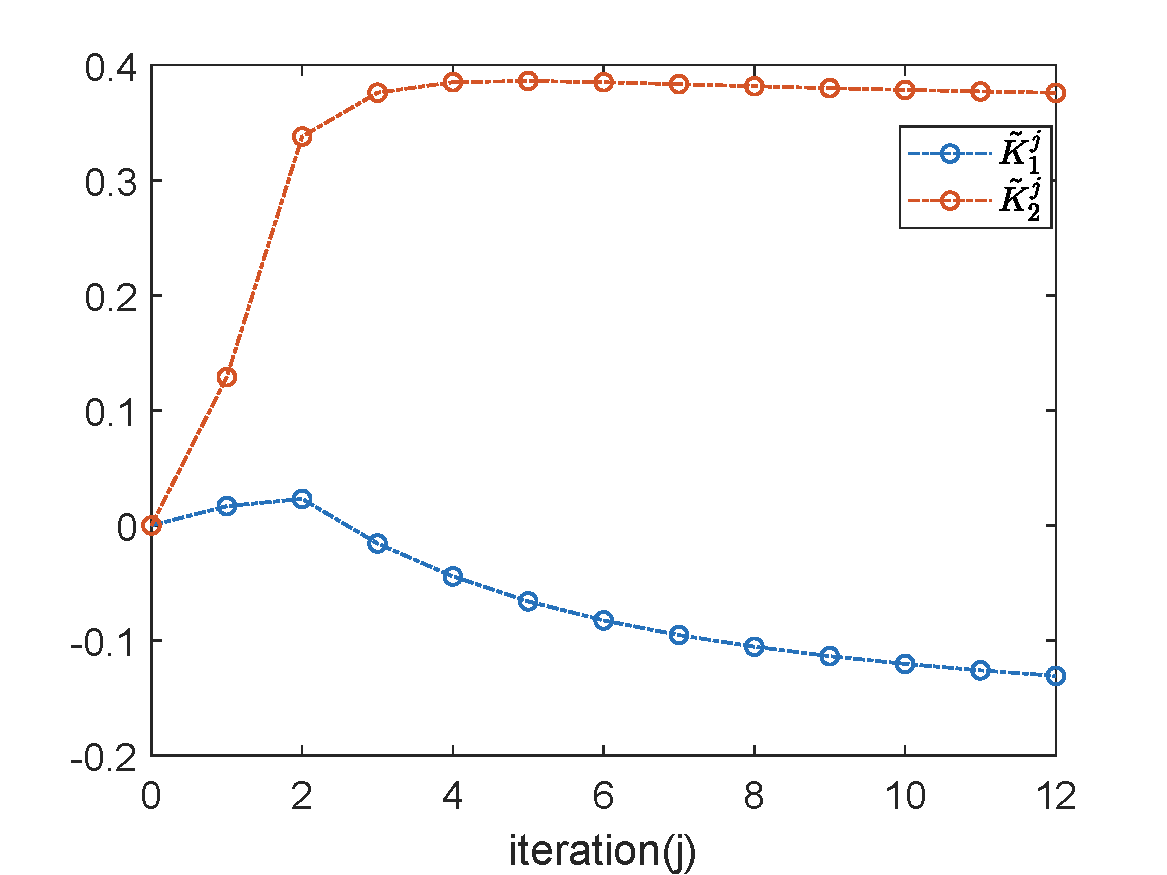}} 
	  \caption{(a). The closed-loop spectral radius $\rho(A-B\tilde{K}^{j})$ obtained by using Algorithm \ref{alg1}; (b). iterations of the matrix $\tilde{K}^{j}$.}
	  \label{fig2}
\end{figure}

To satisfy condition \eqref{26}, we use random probing noise distributed between the intervals $[-1, 1]$ (or a probing noise superimposed by high-frequency sine and cosine functions). When enough data have been collected, the stabilizing control gain $\tilde{K}^{j}$ is obtained by Phases 1 and 2 of the Algorithm \ref{alg1}, where $\alpha_{j+1}=0.4\times\bar{\alpha}$. The iterative update process is shown in Fig. \ref{fig2}. The iteration stops at step 12, i.e., $\tilde{\beta}+\sum_{m=0}^{j}\alpha_{m}\geq1$ is reached. At this time, $\tilde{K}^{j}=[\tilde{K}_{1}^{j},\tilde{K}_{2}^{j}]$ is iterated as
\begin{equation} \label{73}
\begin{aligned}
\tilde{K}^{12}=[   -0.1307  \quad  0.3761]
\end{aligned}
\end{equation}
and $\rho(A-B\tilde{K}^{j})$ is iterated to $0.1959$ at $j=12$.
As shown in Fig. \ref{fig2}(a), the closed-loop spectral radius $\rho(A-B\tilde{K}^{j})$ always converges with an upper bound of $1/(\tilde{\beta}+\sum_{m=0}^{j}\alpha_{m})$, i.e., $\rho(A-B\tilde{K}^{j})<1/(\tilde{\beta}+\sum_{m=0}^{j}\alpha_{m})$. Moreover, Fig.\ref{fig7} (a) exhibits the value of $\alpha_{j}$ at each step, and we verify Theorems \ref{L2} and \ref{the2} by using this figure and Fig. \ref{fig2} (a). Clearly, $\alpha_{j}$ always satisfies \eqref{14} and \eqref{33}.
Let the control gain be $K^{0}=\tilde{K}^{12}$ and run Algorithm \ref{alg1} until $\|P^{i}-P^{i-1}\|<\varepsilon_{1}$, where $\varepsilon_{1}=10^{-5}$.

\begin{figure}
    \centering
	  \subfloat[]{
       \includegraphics[width=0.48\linewidth]{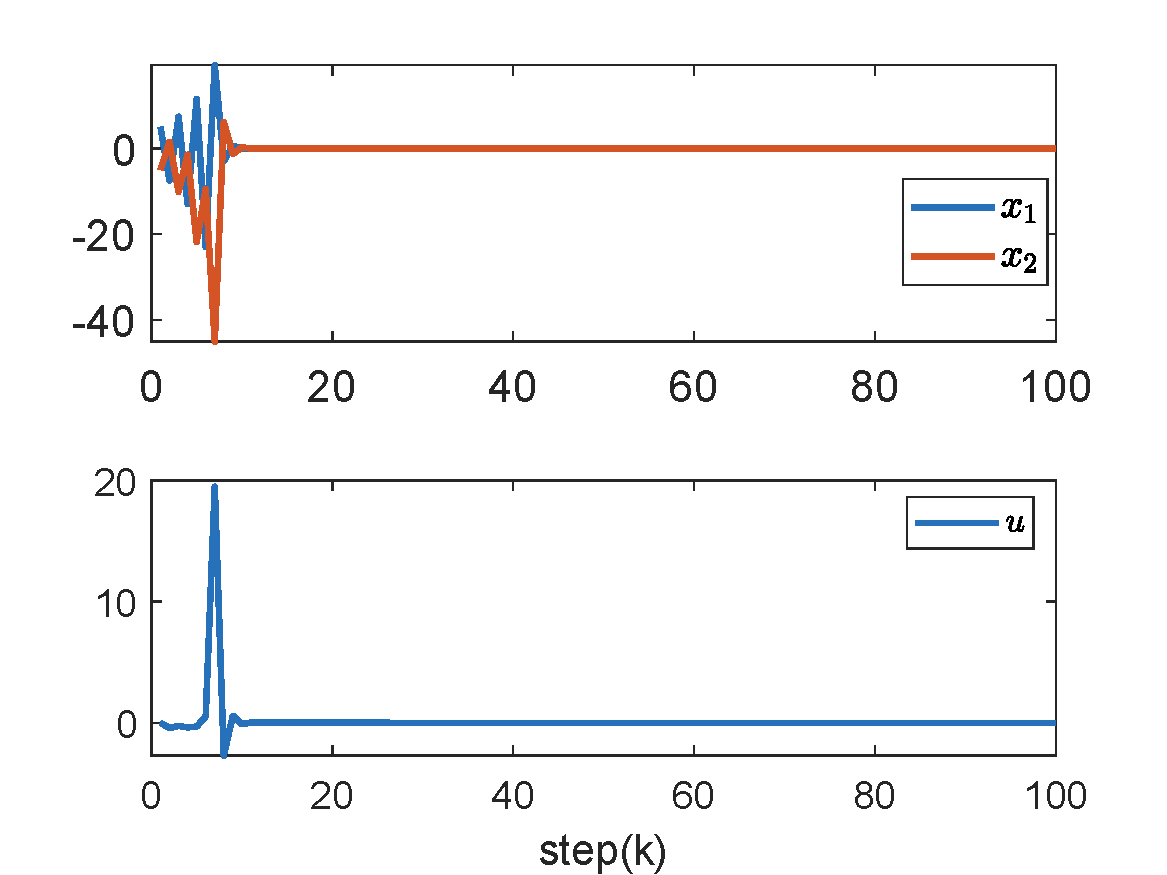}} 
       \hfill
	  \subfloat[]{
        \includegraphics[width=0.48\linewidth]{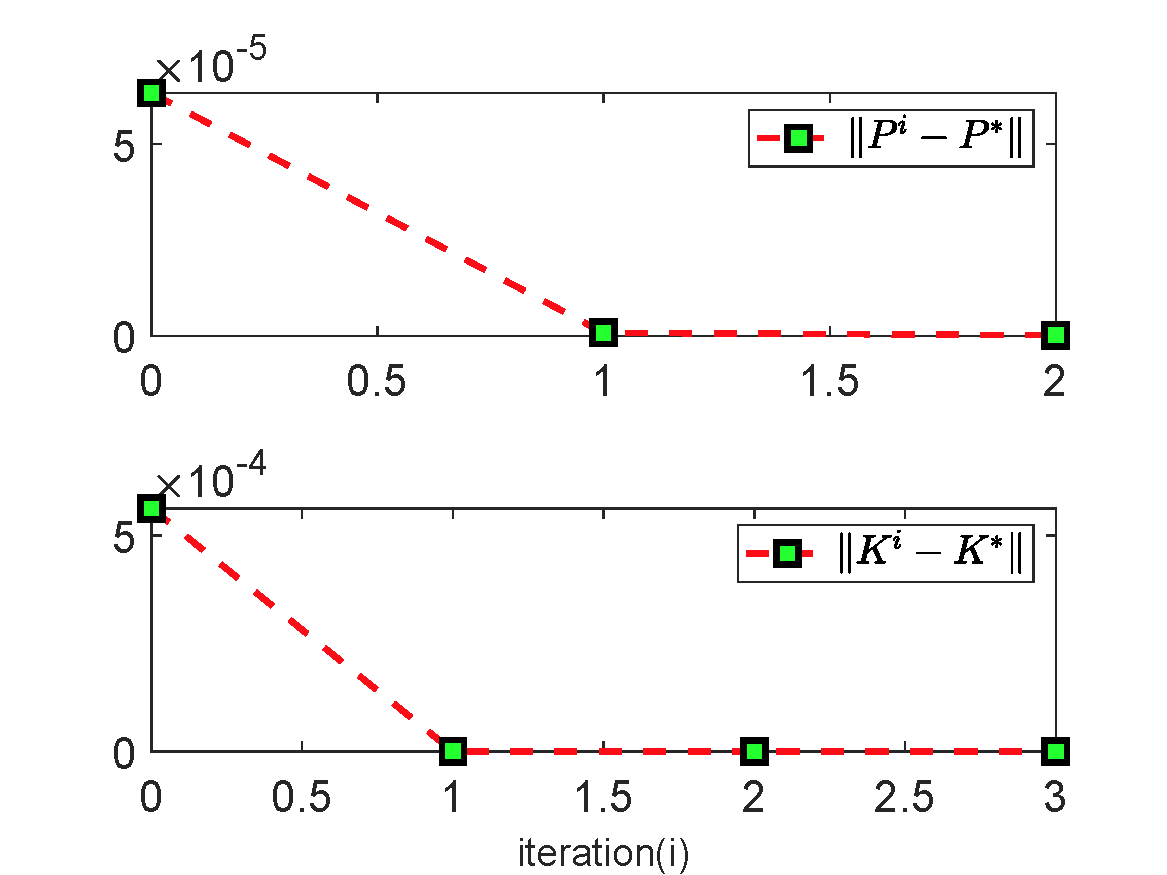}} 
	  \caption{(a). The system states and input obtained by using Algorithm \ref{alg1}; (b). optimal errors of ${P}^{i}$ and ${K}^{i}$  obtained by using Algorithm \ref{alg1}.}
	  \label{fig3}
\end{figure}

\begin{figure}
    \centering
	  \subfloat[]{
       \includegraphics[width=0.48\linewidth]{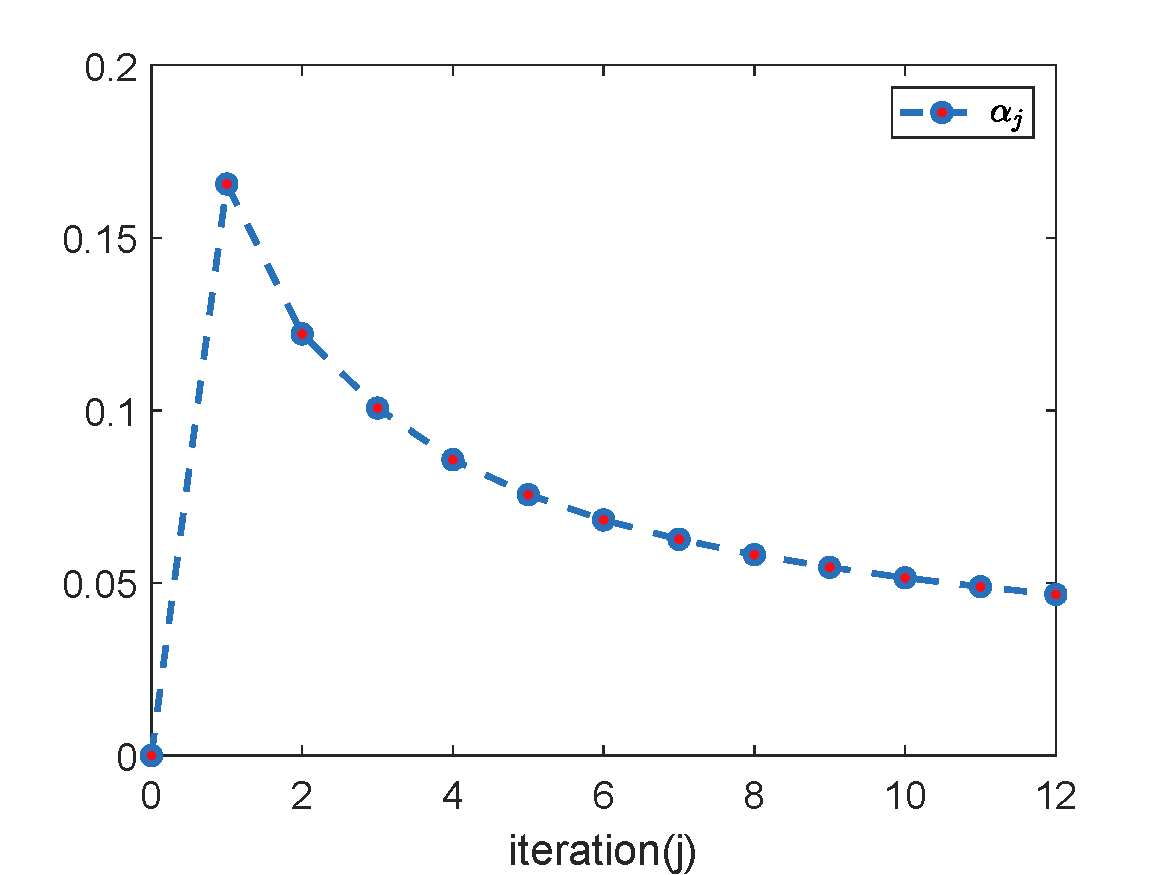}}
       \hfill
	  \subfloat[]{
        \includegraphics[width=0.48\linewidth]{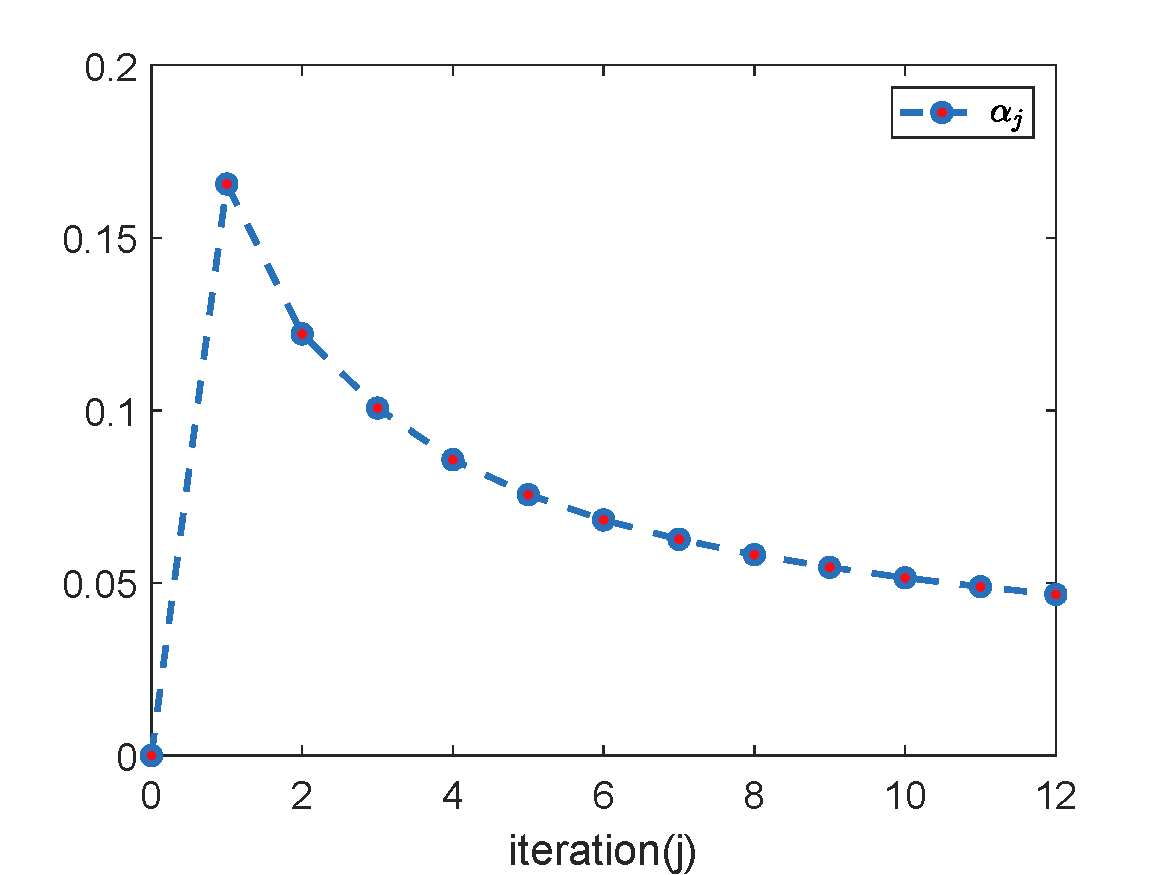}}
	  \caption{(a). The damping coefficient $\alpha_{j}$ obtained by using Algorithm \ref{alg1}; (b). the damping coefficient $\alpha_{j}$ obtained by using Algorithm \ref{alg2}.}
	  \label{fig7}
\end{figure}

After simulation, the results are obtained as in Figs. \ref{fig3}, where Fig. \ref{fig3}(a) shows the states and the input of the system. The system is stabilized under the action of Algorithm \ref{alg1}. From Fig. \ref{fig3}(b), it can be shown that $P^{i}$ and $K^{i}$ can be converged to $P^*$ and $K^*$ quickly by PI. The optimal error is obtained as $\|P^{i}-P^{*}\|=2.1842\times 10^{-8}$.
The final near-optimal solution to ARE \eqref{6} is obtained as
\begin{equation} \label{74}
\begin{aligned}
P^{2}=\left[
\begin{array}{cc}
 27.8194  &  7.5337\\
    7.5337  &  8.7798
\end{array}
\right]
\end{aligned}
\end{equation}
and the near-optimal feedback control gain is
\begin{equation} \label{75}
\begin{aligned}
{K}^{3}=[  -0.1313  \quad  0.3759].
\end{aligned}
\end{equation}
Comparison of the above results with \eqref{71} shows that optimal control is achieved through Algorithm \ref{alg1}.

\subsection{Validation of Off-Policy $\mathcal{Q}$-Learning Algorithm \ref{alg2}}\label{sec.6.2}
\begin{figure}
    \centering
	  \subfloat[]{
       \includegraphics[width=0.48\linewidth]{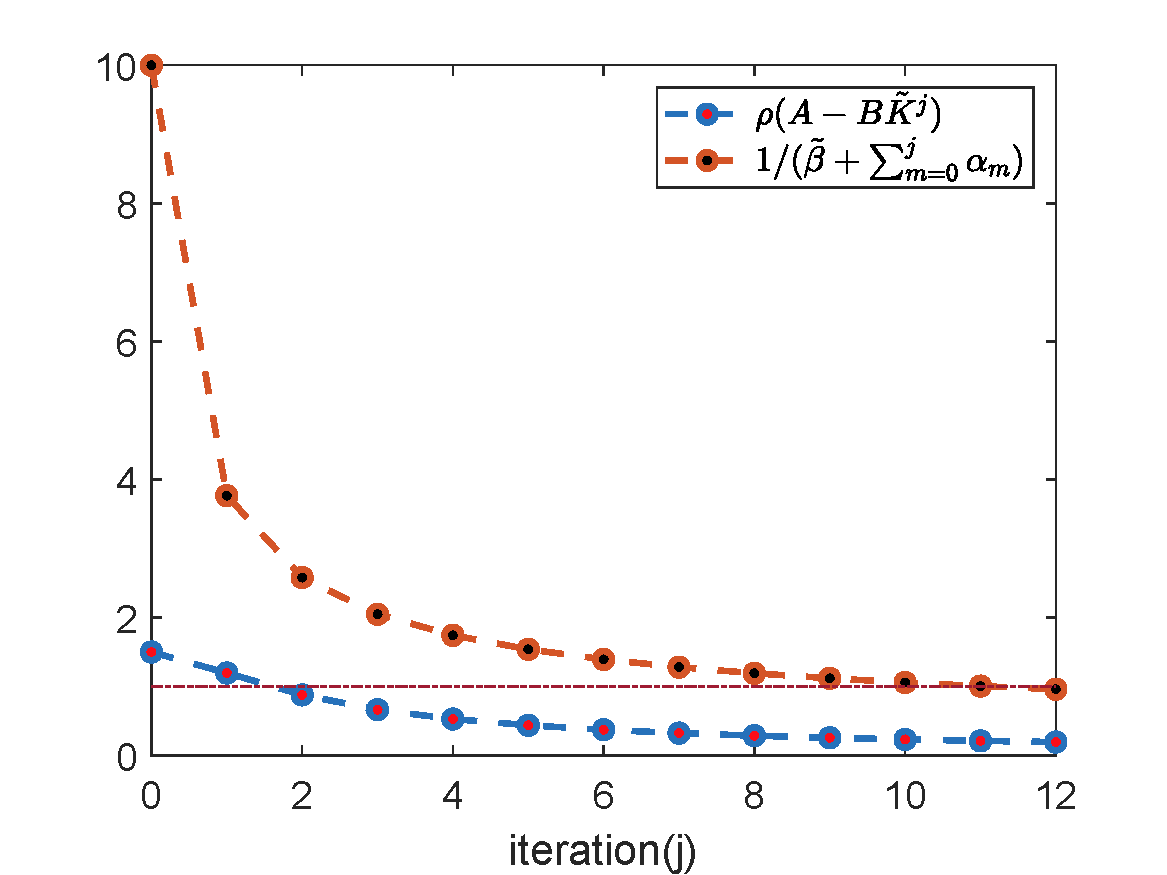}} 
       \hfill
	  \subfloat[]{
        \includegraphics[width=0.48\linewidth]{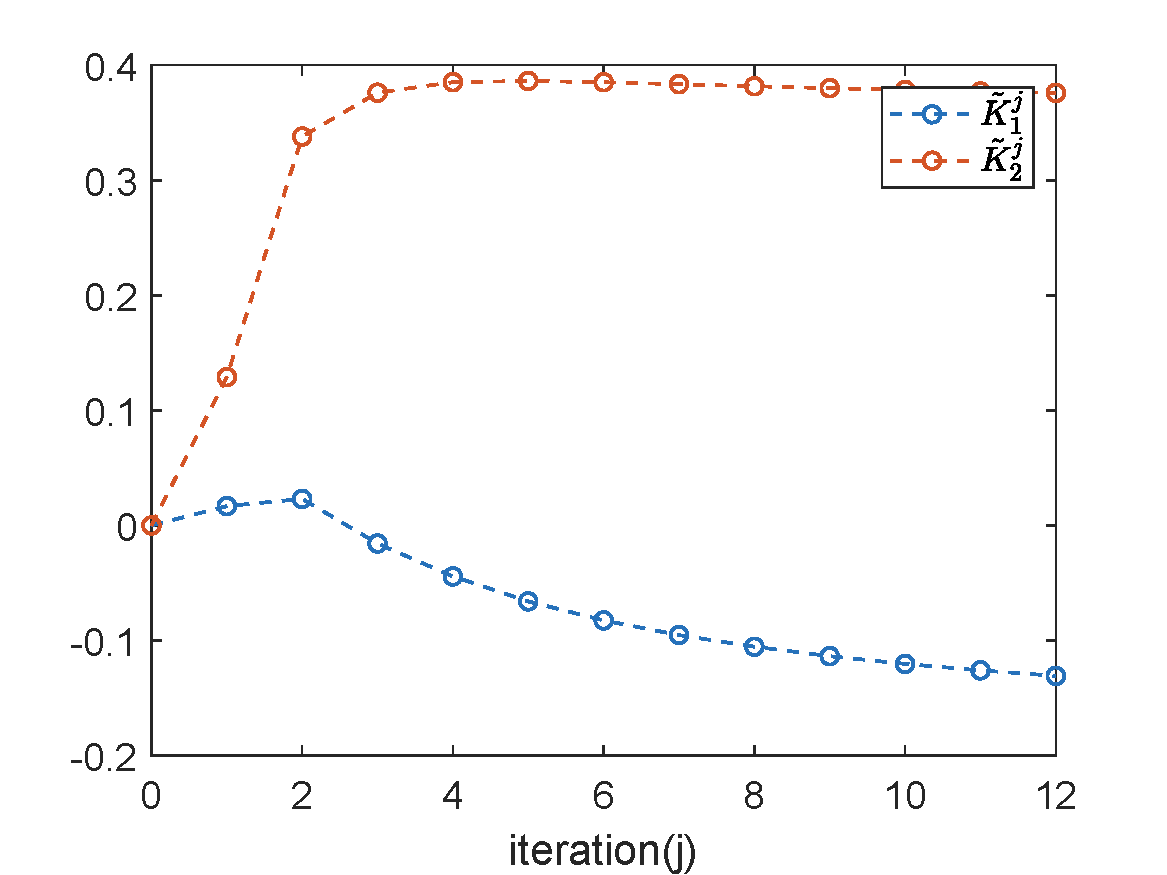}} 
	  \caption{(a). The closed-loop spectral radius $\rho(A-B\tilde{K}^{j})$ obtained by using Algorithm \ref{alg2}; (b). iterations of the matrix $\tilde{K}^{j}$ obtained by using Algorithm \ref{alg2}.}
	  \label{fig4}
\end{figure}
\begin{figure}
    \centering
	  \subfloat[]{
       \includegraphics[width=0.48\linewidth]{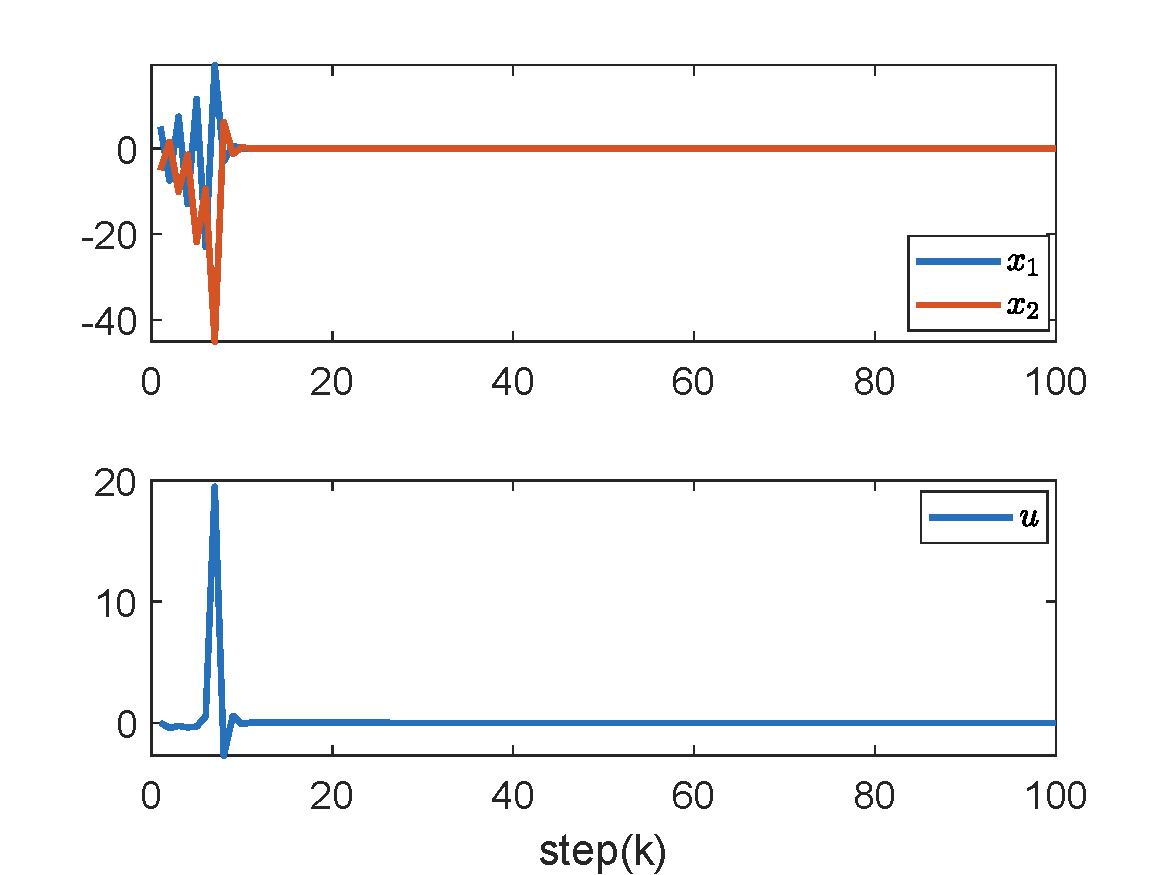}} 
       \hfill
	  \subfloat[]{
        \includegraphics[width=0.48\linewidth]{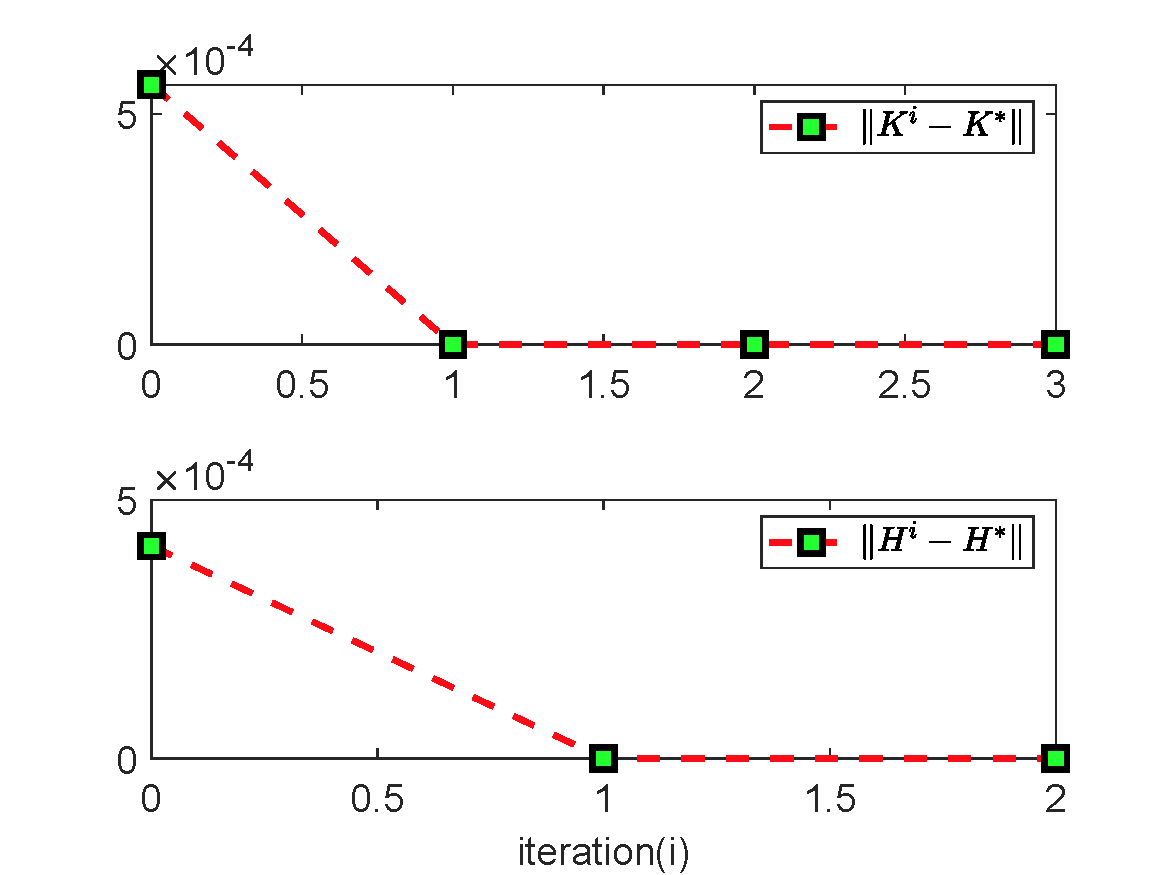}} 
	  \caption{(a). The system states and input obtained by using Algorithm \ref{alg2}; (b). the optimal errors of ${K}^{i}$ and ${H}^{i}$ obtained by using Algorithm \ref{alg2}.}
	  \label{fig5}
\end{figure}

Set $\tilde{\beta}=0.1$, $\alpha_{0}=0.0001$ and $\tilde{K}^{0}=[0,0]$. Choose $\alpha_{j+1}$ according to \eqref{72a} with $a=0.4$. A probing noise identical to Algorithm \ref{alg1} is used to collect data to satisfy \eqref{48}, and subsequently iterates Phases 1 and 2 of Algorithm \eqref{alg2}. The iteration results are shown in Fig. \ref{fig4}, and after 4 steps iterations, a stabilizing control gain is obtained as
\begin{equation} \label{76}
\begin{aligned}
\tilde{K}^{12}=[   -0.1307  \quad  0.3761]
\end{aligned}
\end{equation}
and $\rho(A-B\tilde{K}^{j})$ is iterated to $0.1959$ at $j=12$.
As shown in Fig. \ref{fig4}(a), $\rho(A-B\tilde{K}^{j})$ always converges with an upper bound of $1/(\tilde{\beta}+\sum_{m=0}^{j}\alpha_{m})$, i.e., it satisfies $\rho(A-B\tilde{K}^{j})<1/(\tilde{\beta}+\sum_{m=0}^{j}\alpha_{m})$. Fig. \ref{fig7} (b) shows $\alpha_{j}$ at each step in Algorithm \ref{alg2}. By using this figure and Fig. \ref{fig4} (a), Theorems \ref{L2} and \ref{the2} are verified again. Clearly, $\alpha_{j}$ always satisfies \eqref{14} and \eqref{33} in Algorithm \ref{alg2}.

Let the control gain be $K^{0}=\tilde{K}^{12}$ and run Algorithm \ref{alg2} until $\|H^{i}-H^{i-1}\|<\varepsilon_{2}$, where $\varepsilon_{2}=10^{-5}$. The simulation results are shown in Fig. \ref{fig5}. After obtaining a stabilizing control gain, the solution of the $\mathcal{Q}$-function can be quickly converged by Algorithm \ref{alg2}. The optimal error of the algorithm is finally $\|H^{i}-H^{*}\|=1.4512\times10^{-9}$.
Ultimately, the near-optimal solution to the $\mathcal{Q}$-function \eqref{53} is
\begin{equation} \label{77}
\begin{aligned}
H^{2}=\left[
\begin{array}{ccc}
    30.9728 &  -1.4963  &-24.0202\\
   -1.4963  & 34.6382  & 68.7845\\
  -24.0202 &  68.7845 & 182.9699
\end{array}
\right]
\end{aligned}
\end{equation}
and the near-optimal control gain is
\begin{equation} \label{78}
\begin{aligned}
{K}^{3}=[-0.1313  \quad  0.3759].
\end{aligned}
\end{equation}
Thus, model-free optimal control is realized.

\begin{remark}
In Sections \ref{sec.6.1} and \ref{sec.6.2}, Algorithms \ref{alg1} and \ref{alg2} are verified to be valid, respectively. Note that in both simulations, $\tilde{\beta}$ and $a$ are the same, which leads to the same results (compare Fig. \ref{fig2} and Fig. \ref{fig4}, compare Fig. \ref{fig7}(a) and Fig. \ref{fig7}(b), compare \eqref{73} and \eqref{76}). Thus, Lemma \ref{L4} is verified. The proposed $\mathcal{Q}$-learning algorithm framework proposed is equivalent to the proposed PI framework.
\end{remark}

\subsection{Testing of different damping coefficients $\tilde{\beta}$ and $\alpha_{j+1}$}

\begin{figure}
    \centering
	  \subfloat[]{
       \includegraphics[width=0.48\linewidth]{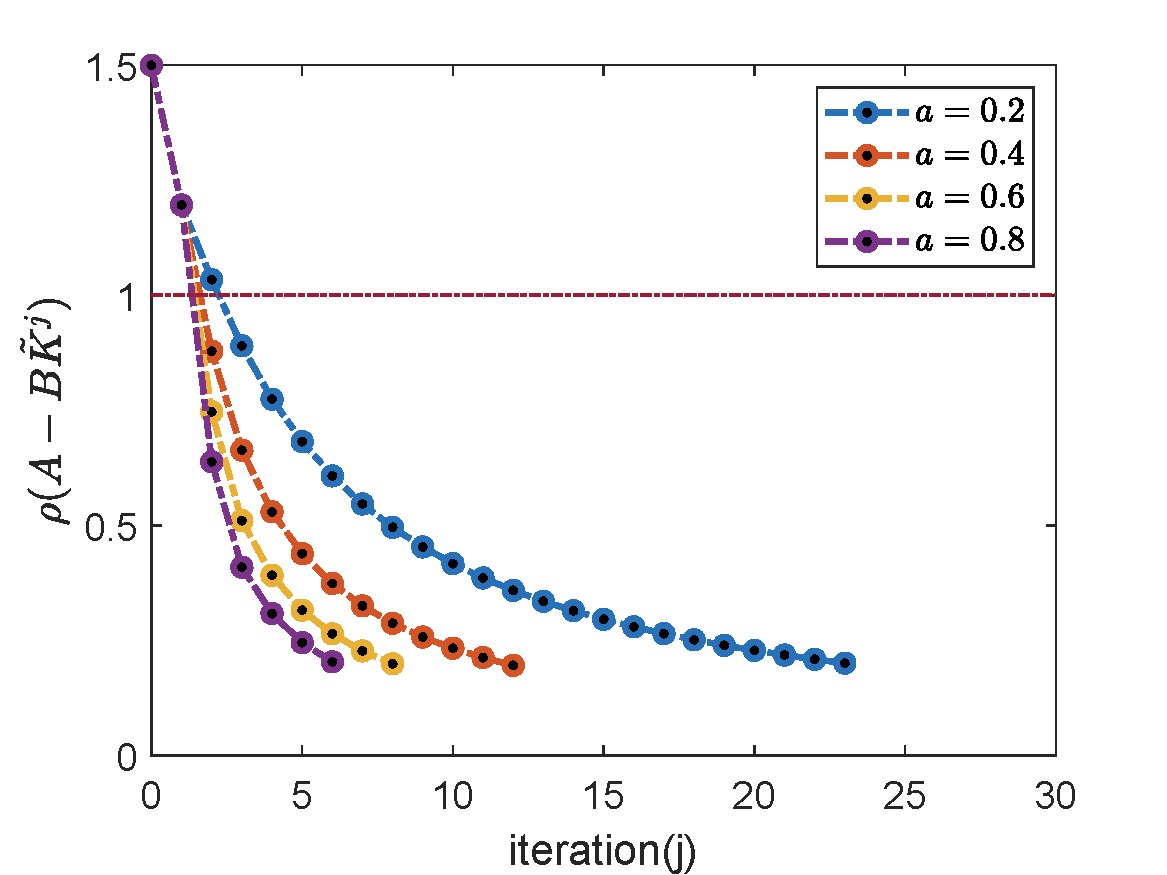}} 
       \hfill
	  \subfloat[]{
        \includegraphics[width=0.48\linewidth]{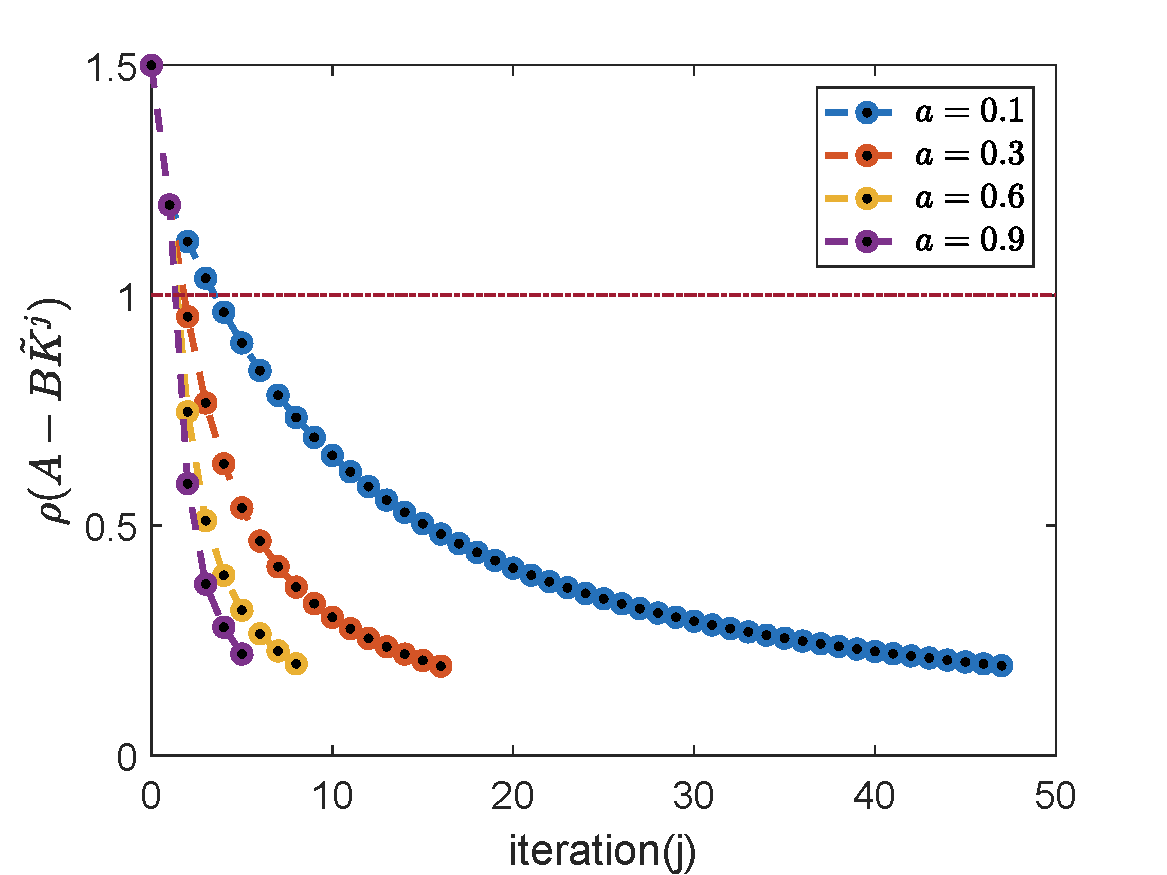}}
	  \caption{(a). The closed-loop spectral radius $\rho(A-B\tilde{K}^{j})$ obtained by using Algorithm \ref{alg1} ($\tilde{\beta}=0.1$); (b). the closed-loop spectral radius $\rho(A-B\tilde{K}^{j})$ obtained by using Algorithm \ref{alg2} ($\tilde{\beta}=0.1$).}
	  \label{fig6}
\end{figure}

To verify the effect of choosing $\alpha_{j+1}$ on Algorithm \ref{alg1} (i.e., the discussion in Remark \ref{rem5}), the result in Fig. \ref{fig6}(a) is obtained by using different ``$a$'' according to \eqref{72a}. The larger coefficient $a$ is set, the larger the $\alpha_{j+1}$ is chosen at each step. As shown in Fig. \ref{fig6}(a), a larger $\alpha_{j+1}$ is chosen, the stabilizing control gain is obtained faster and the closed-loop spectral radius converges faster. Remark \ref{rem5} is verified.

Similar to Algorithm \ref{alg1}, the result in Fig. \ref{fig6}(b) is obtained by using Algorithm \ref{alg2} and different ``$a$'' according to \eqref{72a}. From Fig. \ref{fig6}(b), the conclusion of Remark \ref{rem5} is verified again. As shown in Fig. \ref{fig6}(b), it should be noted that a larger $\alpha_{j+1}$ does not necessarily yield a smaller closed-loop spectral radius, and the size of the $\alpha_{j+1}$ only affects the convergence rate of the closed-loop spectral radius.
\begin{figure}[htbp]
      \centering
      \includegraphics[width=6.6cm,height=4.6cm]{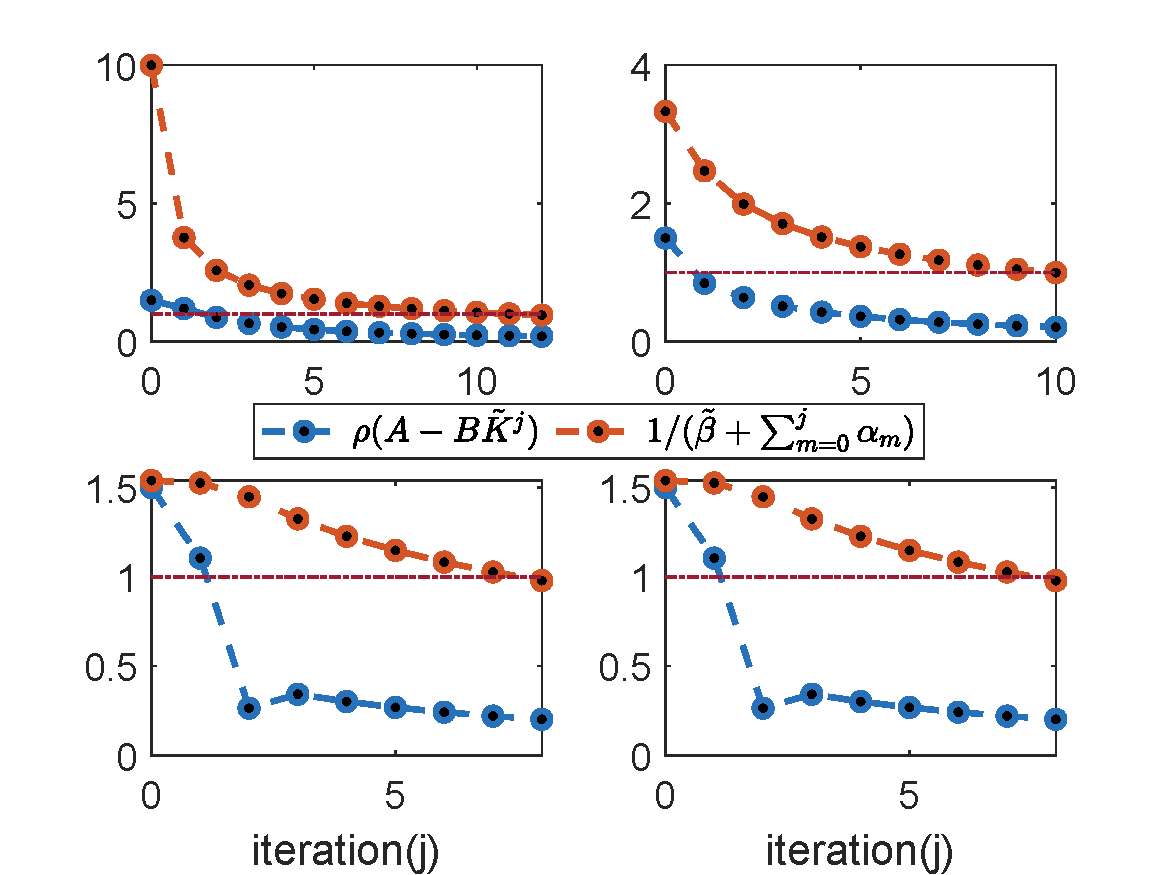}
      \caption{The $\rho(A-B\tilde{K}^{j})$ at different $\tilde{\beta}$ obtained by Algorithm \ref{alg1}: upper left ($\tilde{\beta}=0.1$, $a=0.4$), upper right ($\tilde{\beta}=0.4$, $a=0.4$), lower left ($\tilde{\beta}=0.7$, $a=0.4$) and lower right ($\tilde{\beta}=0.9$, $a=0.4$).}\label{fig14}
\end{figure}
\begin{figure}[htbp]
      \centering
      \includegraphics[width=6.6cm,height=4.6cm]{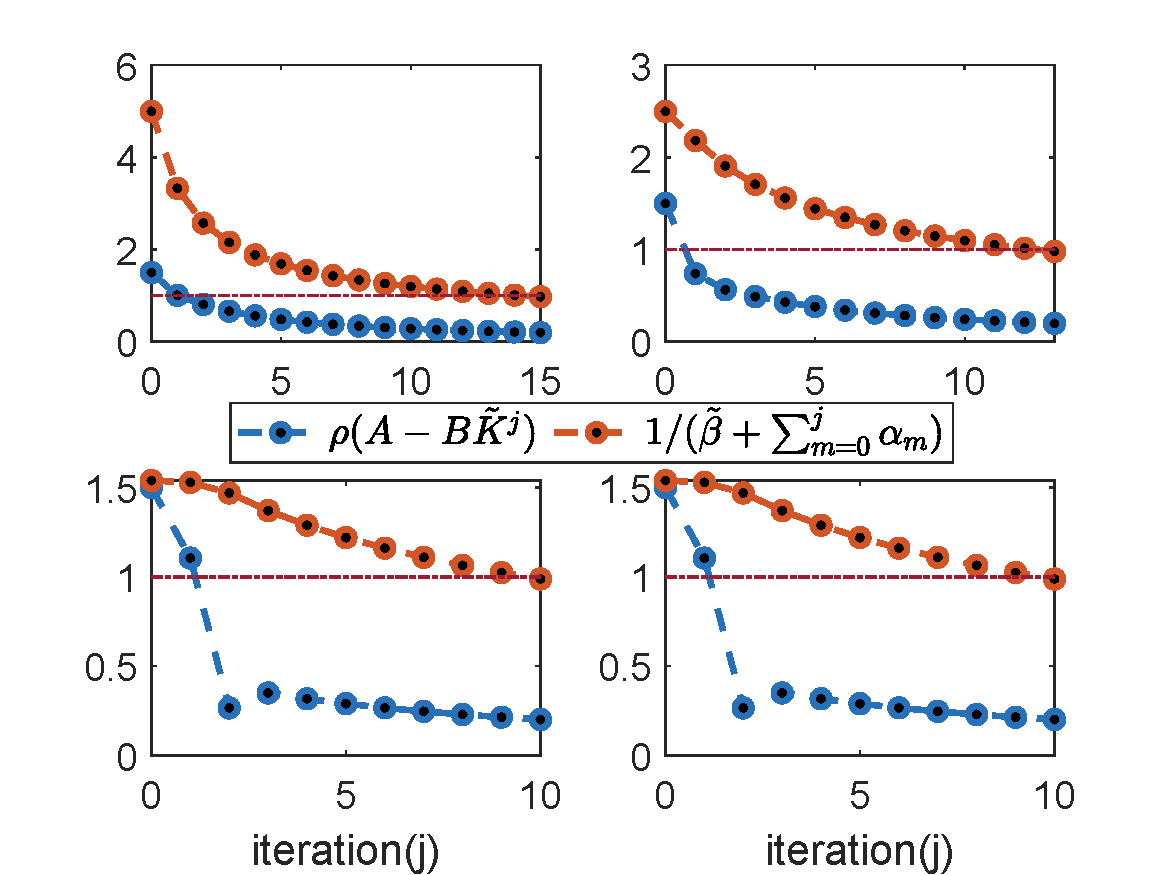}
      \caption{The $\rho(A-B\tilde{K}^{j})$ at different $\tilde{\beta}$ obtained by Algorithm \ref{alg2}: upper left ($\tilde{\beta}=0.2$, $a=0.3$), upper right ($\tilde{\beta}=0.4$, $a=0.3$), lower left ($\tilde{\beta}=0.8$, $a=0.3$) and lower right ($\tilde{\beta}=0.95$, $a=0.3$).}\label{fig15}
\end{figure}

Further, we verify the effect of different $\tilde{\beta}$ on the two algorithms. It should be emphasized that the open-loop spectral radius of system \eqref{70} is $\rho(A)=1.5$. That is, Phase 1 of Algorithms \ref{alg1} and \ref{alg2} is not iterated when $\tilde{\beta}<1/1.5-\alpha_{0}$. When $\tilde{\beta}\geq1/1.5-\alpha_{0}$ is set, the $\tilde{\beta}$ is adjusted to $\tilde{\beta}<1/1.5-\alpha_{0}$ by Phase 1 of Algorithms \ref{alg1} and \ref{alg2}. The validity of Phase 1 of Algorithms \ref{alg1} and \ref{alg2} is verified as shown in the lower left and lower right figures in Figs. \ref{fig14} and \ref{fig15}, respectively. As in Figs. \ref{fig14} and \ref{fig15}, it is also verified that in the interval $\tilde{\beta}\in(0,1/\rho(A)-\alpha_{0})$, setting the initial $\tilde{\beta}$ larger makes $\rho(A-B\tilde{K}^{j})$ converge faster.

\section{Conclusion}\label{section:7}
In this paper, we propose two novel data-based stabilizing optimal control algorithms for DT systems. A stable artificial system is constructed through the damping coefficients, and a stabilizing control policy is obtained by varying the damping coefficients gradually iterating this system to the original closed-loop system. The off-policy iteration algorithm and off-policy $\mathcal{Q}$-learning algorithm are designed to realize model-free optimal control. The convergence of these two algorithms is analyzed. By simulation of an open-loop unstable system, it is verified that the proposed algorithms can not only compute the stabilizing gain and optimal solution, but also more efficient. Notably, model-free PI for DT nonlinear systems has a broader application scenario and is also limited by the initial stabilizing policy, which is the focus of our future work. Moreover, we will endeavor to apply the methods proposed in this paper to real scenarios.

\appendices
\section{Proof of Lemma \ref{rank1}}\label{app1}
Suppose there exists a nonzero solution $Y=[Y_{p}^{T},Y_{1}^{T},Y_{2}^{T}]^{T}$ such that $\psi^{j}Y=0$, where $Y_{p}=vecs(\tilde{Y}_{p})$, $Y_{1}=vec(\tilde{Y}_{1})$ and $Y_{2}=vecs(\tilde{Y}_{2})$ with $\tilde{Y}_{p}=\tilde{Y}_{p}^{T}$ and $\tilde{Y}_{2}=\tilde{Y}_{2}^{T}$.  According to $\psi^{j}Y=0$ and \eqref{21}, it is easy to obtain
\begin{equation} \label{AP1}
\begin{aligned}
0=&\Xi_{2}vecs(\chi_{p})+\Xi_{3}vec(\chi_{1})+\Xi_{4}vecs(\chi_{2})
\end{aligned}
\end{equation}
where $\chi_{p}=\tilde{A}^{jT}\tilde{Y}_{p}\tilde{A}^{j}-\tilde{Y}_{p}+2\tilde{K}^{jT}(A^{T}\tilde{Y}_{p}B-\tilde{Y}_{1})-\tilde{K}^{jT}(B^{T}\tilde{Y}_{p}B-\tilde{Y}_{2})\tilde{K}^{j}$, $\chi_{1}=2(A^{T}\tilde{Y}_{p}B-\tilde{Y}_{1})$ and $\chi_{2}=B^{T}\tilde{Y}_{p}B-\tilde{Y}_{2}$ with $\tilde{A}^{j}$ defined in \eqref{18}.

Clearly, the matrix $[\Xi_{2},\Xi_{3},\Xi_{4}]$ is  full-column rank if \eqref{26} is satisfied. Then \eqref{AP1} has the unique solution as $[vecs(\chi_{p})^{T},vec(\chi_{1})^{T},vecs(\chi_{2})^{T}]^{T}=0$. Further, one can get $\tilde{A}^{jT}\tilde{Y}_{p}\tilde{A}^{j}-\tilde{Y}_{p}=0$, where $\tilde{A}^{j}$ is Schur. Therefore, $\tilde{Y}_{p}$ must be zero, and then $Y_{p}$ must be zero. Moreover, there exist $\tilde{Y}_{1}=A^{T}\tilde{Y}_{p}B=0$ and $\tilde{Y}_{2}=B^{T}\tilde{Y}_{p}B=0$. Finally, we can get $Y = 0$. This clearly contradicts the non-zero assumption. Therefore, $\psi^{j}$ is full-column rank .

\section{Proof of Lemma \ref{lem4}}\label{app2}
Since $\mathcal{M}=\gamma_{j}\left[\begin{array}{c}
I_{n_x}\\
-\tilde{K}\\
\end{array}
\right][A\quad B]$, there exists an invertible matrix ${\Lambda}$ such that
\begin{equation} \label{AP2}
\begin{aligned}
{\Lambda}^{-1}\mathcal{M}{\Lambda}=\gamma_{j}\left[
\begin{array}{cc}
 A-B\tilde{K} & B \\
0 & 0\\
\end{array}
\right]
\end{aligned}
\end{equation}
It is easy to obtain
\begin{equation} \label{AP3}
\begin{aligned}
{\Lambda}=\gamma_{j}\left[
\begin{array}{cc}
 I_{n_x} &  I_{n_u} \\
-\tilde{K} & 0\\
\end{array}
\right].
\end{aligned}
\end{equation}
From \eqref{AP2}, $\mathcal{M}$ is stable if and only if $\gamma(A-B\tilde{K})$ is stable. Then, the Lyapunov equation \eqref{42c} has a unique positive definite solution $\tilde{H}$.

\bibliographystyle{IEEEtran}
\bibliography{Reference}

\end{document}